%% file: main.tex
\documentclass[11pt]{article}
\usepackage{graphicx}
\usepackage{float}
\usepackage{natbib} 
\usepackage{url} 
\usepackage{amsmath} 
\usepackage{amsfonts}
\usepackage{amssymb} 
\usepackage{lineno}
\usepackage{setspace}
\usepackage{lscape} 
\usepackage{float} 
\usepackage{breakcites}
\usepackage{pdfpages}
\usepackage{bm}
\usepackage{xcolor}
\usepackage{hyperref}
\usepackage{setspace}
\usepackage{amsthm}
\usepackage{booktabs}
\usepackage{subcaption}

\usepackage[ruled,vlined]{algorithm2e}

\graphicspath{{Figures/}}

\usepackage{mathtools}

\newcommand{\interior}[1]{%
 {\kern0pt#1}^{\mathrm{o}}%
}
\usepackage [english]{babel}
\usepackage [autostyle, english = american]{csquotes}
\MakeOuterQuote{"}

\usepackage{caption}

\usepackage[makeroom]{cancel}

\newcommand{\blind}{0}

\usepackage[letterpaper,margin=1in]{geometry}

\usepackage{lineno}
\begin{document}

\def\spacingset#1{\renewcommand{\baselinestretch}%
{#1}\small\normalsize} \spacingset{1}

\if0\blind
{

\title{\bf Physics-Constrained Modeling and Inference of the Equation of State for Dissociative Materials}

\author{
Jolypich Pek$^{1}$,
Ben Seiyon Lee$^{1,*}$,
Jason Bernstein$^{2}$,
and Philip Myint$^{2}$\\[0.2cm]
$^{1}$Department of Statistics, George Mason University, Fairfax, Virginia, USA\\
$^{2}$Lawrence Livermore National Laboratory, Livermore, California, USA\\[0.1cm]
$^{*}$Corresponding author: \texttt{slee287@gmu.edu}
}

\maketitle
}
 \fi

\if1\blind
{
  \bigskip
  \bigskip
  \bigskip
  \begin{center}
    {\LARGE\bf Physics-Constrained Modeling and Inference of the Equation of State for Dissociative Materials}
\end{center}
  \medskip
} \fi

\thispagestyle{empty}
\begin{abstract}
Equations of state (EOSs) describe how a material's internal energy and pressure change with temperature and density. They are inputs to hydrodynamic simulations in materials science and planetary science. We model the ion-thermal free-energy term for dissociative materials, where molecular bonds break apart to form new species under extreme conditions. This term is governed by two latent physical processes, a molar-mass surface and a Debye-temperature curve, neither of which is observed directly. Instead, both must be recovered from internal energy and pressure, which are derivatives of the free energy, subject to thermodynamic constraints. Existing EOS tables are typically built through material-specific hand-tuning, a process that is tedious, does not scale across materials or conditions, and offers no guarantee of thermodynamic consistency. We propose a partitioned statistical framework that infers molar mass and Debye temperature directly from EOS data, guaranteeing thermodynamic consistency by construction. Local molar-mass behavior is captured with polynomial basis functions across domain partitions, while the Debye temperature is modeled with an I-spline basis. In simulations and a carbon dioxide case study, the method accurately recovers the underlying physical quantities, improves prediction of held-out energy and pressure data, and provides uncertainty quantification while scaling with EOS table size.

\end{abstract}

\noindent%
\vfill

\clearpage
\setcounter{page}{1}
\newpage
\spacingset{2} 

\section{Introduction}
\label{sec:intro}
An equation of state (EOS) describes the relationship among a material's thermodynamic variables such as pressure, internal energy, entropy, temperature, and density \citep{landau1980statistical,toro2009equations}. It is commonly tabulated over density and temperature for use in hydrodynamic simulations foundational to thermodynamics, fluid dynamics, planetary science, and materials science \citep{lyon1992alamos,toro2009equations, drake2010high}. In high-energy-density science, such EOS tables are essential for predicting material response under extreme compression, heating, and phase change \citep{atzeni2004physics, lindl2004physics}, for interpreting experiments, guiding target design, and quantifying uncertainties in large-scale simulations \citep{drake2010high, kritcher2022design}.

A common EOS formulation decomposes the material's free energy into cold, ion-thermal, and electron-thermal contributions \citep{more1988new}. We focus on the ion-thermal term, which is governed by two physical latent processes, a molar mass and the Debye temperature. The molar mass varies with both density and temperature and reflects atomic or molecular composition, and the Debye temperature which varies density and reflects how that composition responds thermally under compression and heating \citep{benedict2014multiphase}. EOS modeling can be challenging because neither the molar-mass surface nor the Debye-temperature curve is observed directly \citep{more1988new,lindquist2022uncertainty}. Instead, the observed data are internal energy and pressure, which are functions of the derivatives of the free energy which the two latent quantities jointly determine. Recovering these processes therefore requires solving an inverse problem constrained by thermodynamic consistency. That is, it must obey the exact mathematical relationships thermodynamics places on these derivatives.



Existing inference methods rely largely on manual tuning to recover these latent processes: parameters are estimated from static compression, shock and ramp compression experiments, and first-principles calculations \citep{more1988new,benedict2014multiphase,rehn2021multiphase,myint2023fluid}, then hand-tuned iteratively using visual heuristics until the fit looks acceptable \citep{myint2023fluid}. More automated alternatives exist: \citet{ali2020development} combined deterministic optimization with Monte Carlo resampling to construct a copper EOS with propagated uncertainty; related deterministic and swarm-intelligence strategies have since been applied to multiphase carbon and other materials \citep{cox2015swarm,cox2018generating}; Bayesian calibration with uncertainty quantification has been applied to multi-phase carbon \citep{lindquist2022uncertainty} and magnesium \citep{lindquist2023uncertainty} EOS models, and extended to jointly infer EOS parameters alongside unknown data-error terms for the unreacted high explosive PBX 9501 \citep{schmid2025simultaneous}; and physics-informed Gaussian processes model free energy, internal energy, and pressure jointly so that thermodynamic consistency holds by construction \citep{sharma2024learning}. Across these approaches, the Debye temperature is an estimable quantity and the molar mass is fixed, since none of these studies considers dissociation.

Dissociation poses a distinct additional challenge for this inference problem. As a material is compressed and heated, molecular bonds break and new species form; for example, carbon dioxide (CO$_2$) dissociates into a mixture of CO$_2$, CO, C, O$_2$, and other species at high density and temperature \citep{ross1985matter}. Since molar mass depends on composition, it varies as dissociation changes the species present in the mixture \citep{wu2019equation}. Molar mass must therefore be treated as an unknown function of density and temperature rather than a fixed constant, and neglecting this shifts predicted pressure and heat capacity relative to models that allow composition to vary \citep{chen2006self,chen2008dissociation,wu2019equation}. Yet most EOS studies do not consider dissociation at all, as they treat single-element or fixed-composition materials. When dissociation is included, composition is typically supplied as a separately tabulated correction from a standalone chemical-equilibrium or ab initio calculation, rather than inferred jointly with the free-energy model \citep{maerzke2019equations}. \citet{wu2019equation} employ hand-tuning and visual heuristics to specify parameters of a simplified dissociation model, which is then used to construct a density- and temperature-dependent average molar mass surface. To the best of our knowledge, no existing study uses an automated process to jointly infer a varying molar mass with the free energy from heterogeneous data while quantifying uncertainty throughout.



A natural first approach is to model the molar-mass surface as a transformation of a global basis expansion or a Gaussian process (GP). A global basis expansion must represent sharp and smooth features with the same basis, so it tends to oversmooth sharp transitions and oscillate in smooth regions; we use it for comparison in Sections~\ref{sec:simulation} and~\ref{sec:casestudy}. Standard GPs use stationary, isotropic kernels, which impose one correlation structure across the density--temperature domain and thus poorly suit a heterogeneous latent surface. Non-stationary kernels \citep{sampson1992nonparametric,paciorek2006spatial,schmidt2020flexible} relax this at added cost, and exact GPs scale as $O(n^3)$ \citep{rasmussen2003gaussian}; approximations such as Vecchia, nearest-neighbor GPs, and SPDEs \citep{vecchia1988estimation,datta2016hierarchical,lindgren2011explicit} reduce this cost, but all $n$ points of the surface must still be estimated. GPs also do not by themselves enforce nonlinear constraints such as positivity of the molar mass. \citet{sharma2024learning} build thermodynamic constraints into a GP on the free energy, but these are implemented on a single-phase diamond with fixed composition; hence, this does not account for dissociation and latent molar mass is not inferred. These limitations motivate our approach, which adds non-stationary components to capture the heterogeneity of the latent molar-mass surface.

We propose a joint, partitioned inferential framework that directly addresses this methodological and scientific gap, estimation of the latent processes while accounting for dissociation. Partitioning the domain handles the multiscale structure directly, letting each region use basis functions suited to its own behavior. To maintain flexibility, we infer the molar-mass surface and Debye-temperature curve via local polynomial expansions and monotone I-splines. Thermodynamic consistency holds by construction, since internal energy and pressure are both derived from a single Cowan free energy with the two fields modeled jointly. Domain partitioning enables scalable inference while capturing localized nonstationary dependence, and uncertainty in both the latent surfaces and resulting EOS predictions is quantified via parametric bootstrapping. We demonstrate the approach on a simulation study and a CO$_2$ EOS case study, enabling investigation of the thermodynamic behavior of dissociating materials at pressures and temperatures with greater confidence.

The remainder of this paper is organized as follows. Section~\ref{sec:EOS} formulates the EOS problem, including the free-energy decomposition and the roles of the molar-mass surface and Debye temperature. Section~\ref{sec:method} develops the partitioned framework for jointly estimating these surfaces. Section~\ref{sec:UQ} describes uncertainty quantification via parametric bootstrapping. Sections~\ref{sec:simulation} and~\ref{sec:casestudy} evaluate the method through a simulation study and the CO$_2$ data application. Section~\ref{sec:discussion} concludes with a summary and directions for future work.

\section{Equations of State}
\label{sec:EOS}


In this study, we focus on EOS models for dissociating materials, using CO$_2$ as a test case. Molecular CO$_2$ is a simple polyatomic chemical species that dissociates at high pressure and temperature and requires EOS methodology that incorporates chemical-equilibrium effects \citep{wu2019equation,chen2008dissociation}. Developing EOS models that account for chemical-equilibrium and dissociation effects is part of a broader push toward wide-ranging EOS models for materials of interest in high-pressure science, including metals \citep{wu2023wide} and generally nonmetallic systems like helium \citep{myint2023fluid}. The high-pressure, high-temperature behavior of CO$_2$ plays an important role in explosive detonations \citep{van1987properties}, Earth's deep carbon cycle \citep{seto2008fate}, and the composition of planetary atmospheres such as Venus \citep{stacey2005high}.

\subsection{Free-Energy Formulation of the Ion-Thermal EOS}
\label{sec:freeenergy}
\begin{figure}[!htb]
    \centering    \includegraphics[width=0.9\textwidth]{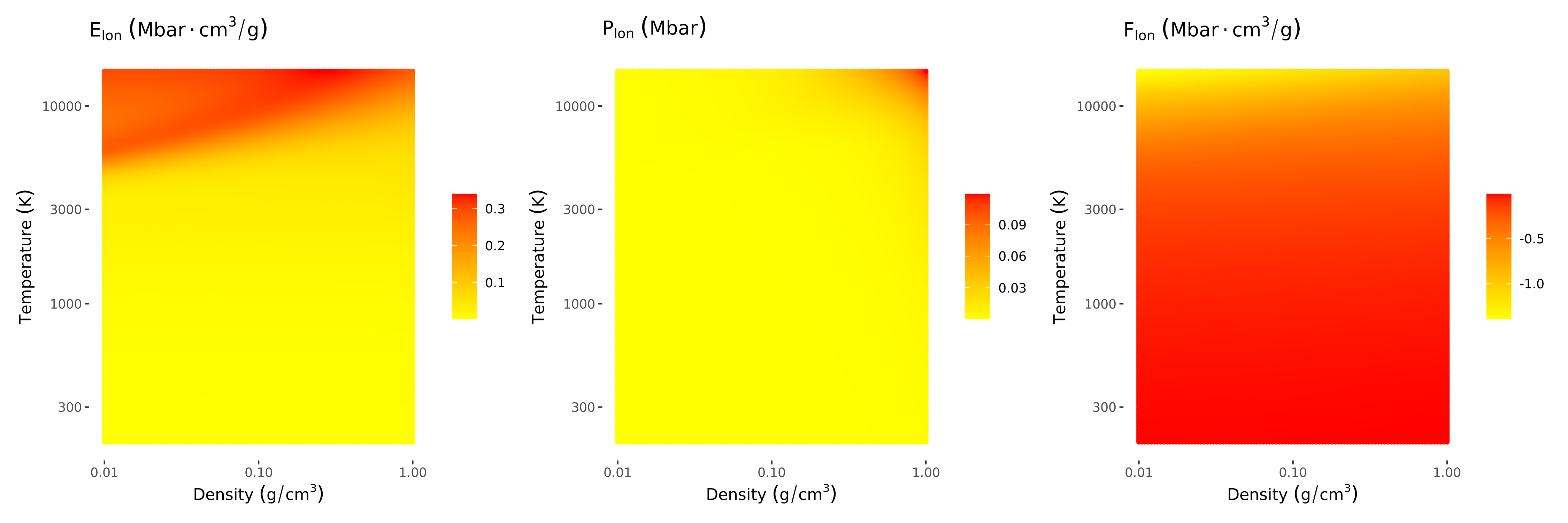}
    \caption{Ion-thermal contributions to the internal energy ($E_{\mathrm{ion}}$), pressure ($P_{\mathrm{ion}}$), and free energy ($F_{\mathrm{ion}}$) as functions of density and temperature for CO$_2$.}
    \label{fig:ion_panel}
\end{figure}

EOS tables consist of thermodynamic quantities and their corresponding temperatures and densities. EOS is formulated using the Helmholtz free energy, decomposed into cold, ionic, and electron contributions \citep{more1988new} $F(\rho,T) = F_{\mathrm{cold}}(\rho) + F_{\mathrm{ion}}(\rho,T) + F_{\mathrm{electron}}(\rho,T)$. We focus on $F_{\mathrm{ion}}(\rho,T)$, that accounts for the thermal motion of the ions, which are the atomic nuclei of the dissociated species. The corresponding ionic thermodynamic quantities are:
\begin{align}
    S_{\mathrm{ion}}(\rho,T) &= -\left(\frac{\partial F_{\mathrm{ion}}}{\partial T}\right)_\rho, \label{eq:Entropy}\\
    E_{\mathrm{ion}}(\rho,T) &= F_{\mathrm{ion}}(\rho,T) + T S_{\mathrm{ion}}(\rho,T), \label{eq:InternalEnergy}\\
    P_{\mathrm{ion}}(\rho,T) &= \rho^2 \left(\frac{\partial F_{\mathrm{ion}}}{\partial \rho}\right)_T. \label{eq:Pressure}
\end{align}
Here, $\rho$ is the density and $T$ is the temperature, and $S_{\mathrm{ion}}$, $E_{\mathrm{ion}}$, and $P_{\mathrm{ion}}$ denote the ionic entropy, internal energy, and pressure, respectively. The total internal energy and pressure inherit the cold, ionic, and electron decomposition of $F$ (see Supplementary Material). For the CO$_2$ case study, the three thermodynamic surfaces are provided in Figure~\ref{fig:ion_panel}.

We account for dissociation by allowing the molar mass $M = M(\rho,T)$ to be a density- and temperature-dependent function, rather than a constant. This provides a simple way to represent changes in composition without having to explicitly track the different species that may form as a result of dissociation. This molar mass function $M(\rho,T)$ and Debye temperature $\Theta(\rho)$ are parameters in an underlying free energy $F_{\mathrm{ion}} = F_{\mathrm{ion}}(\rho,T)$, for which we adopt a Cowan-type model \citep{more1988new},
\begin{equation}
    F_{\mathrm{ion}} = F_{\mathrm{Cowan}}(\rho,T)= \frac{RT}{M(\rho,T)}\left[-\frac{11}{2}+\frac{9}{2}w(\rho,T)^{1/3}+\frac{3}{2}\ln\left(\frac{\chi(\rho,T)^2}{w(\rho,T)}\right)\right].
    \label{eq:Cowan}
\end{equation}
where $R$ is the molar gas constant, expressed in units compatible with those of $M$, $\chi(\rho,T) = \Theta(\rho)/T$, and $w(\rho,T) = T_{\mathrm{melt}}(\rho)/T$, with $\Theta(\rho)$ the Debye temperature and $T_{\mathrm{melt}}(\rho) = \alpha\,\rho^{-2/3}\Theta(\rho)^2$ the melting temperature from the Lindemann criterion \citep{lindemann1910calculation}, $\alpha>0$ being a dimensionless material-specific constant. For $\mathrm{CO}_2$, $M(\rho,T)$ varies between the molecular upper bound $M_{\max}=44\,\mathrm{g\,mol^{-1}}$ and the fully dissociated lower bound $M_{\min}=44/3\,\mathrm{g\,mol^{-1}}$. 

\subsection{Inverse Problem}
\label{sec:inverse_problem}
Our data source is the CO$_2$ EOS table from \citet{wu2019equation}, itself constructed from compiling simulation and experimental data. We infer $M$ and $\Theta$ against this table, rather than against the sparse underlying simulation and experimental data directly. A table gives dense, self-consistent coverage over the full density--temperature domain, unlike the sparse and irregularly sampled coverage typical of raw simulation and experimental sources. This lets us directly assess the model's ability to recover the molar-mass and Debye-temperature surfaces and to quantify uncertainty, before extending the framework to sparser, more heterogeneous data. Related work has trained neural network surrogates on compiled EOS tables \citep{mentzer2023neural}.

This sets up an inverse problem where the observations are the tabulated ion-thermal internal energy and pressure, $\left(E_{\mathrm{ion}}^{\mathrm{obs}}, P_{\mathrm{ion}}^{\mathrm{obs}}\right)$, on the $(\rho,T)$ grid. The CO$_2$ table is sampled uniformly in $x_i=\log_{10}\rho_i$ and $y_i=\log_{10}T_i$, so all estimation is carried out in these log-coordinates. From Section~\ref{sec:method} onward we write $x=\log_{10}\rho$ and $y=\log_{10}T$ and express the estimated surfaces as $M(x,y)$ and $\Theta(x)$, while the free-energy relations above remain in the physical variables $(\rho,T)$.

The estimable processes are $\Theta(x)$ and $M(x,y)$, which are inputs to the Cowan free energy \eqref{eq:Cowan}. Given $\Theta$ and $M$, the free energy $F_{\mathrm{Cowan}}$ predicts $E_{\mathrm{ion}}$ and $P_{\mathrm{ion}}$ through the thermodynamic derivatives \eqref{eq:InternalEnergy} and \eqref{eq:Pressure}. We treat the tabulated values as noisy observations of these predictions,
\begin{equation}
\label{eq:obs_model}
E_{\mathrm{ion},i}^{\mathrm{obs}} = E_{\mathrm{ion}}(\rho_i,T_i;\Theta,M) + \varepsilon^{E}_i,
\qquad
P_{\mathrm{ion},i}^{\mathrm{obs}} = P_{\mathrm{ion}}(\rho_i,T_i;\Theta,M) + \varepsilon^{P}_i,
\end{equation}
where the errors are mean-zero, heteroskedastic and have standard deviations that scale with the local magnitude of each response (the exact form is given in Section  \ref{par:loss}, where it arises naturally from the loss function used for estimation). We assume that the errors are independent across grid points and that $\varepsilon_i^E$ and $\varepsilon_i^P$ are independent at each grid point. Estimating $\Theta(x)$ and $M(x,y)$ is thus an inverse problem: we recover the two latent processes $\Theta(x)$ and $M(x,y)$ that best reproduce the observed $\left(E_{\mathrm{ion}}^{\mathrm{obs}}, P_{\mathrm{ion}}^{\mathrm{obs}}\right)$ across the density--temperature domain. Jointly modeling $E_{\mathrm{ion}}$ and $P_{\mathrm{ion}}$ from the same free-energy representation preserves thermodynamic consistency across the fitted EOS surface. Without further restrictions, however, $\Theta(x)$ and $M(x,y)$ are not separately identifiable from $E_{\mathrm{ion}}$ and $P_{\mathrm{ion}}$ alone, since a change in $\Theta$ can be offset by rescaling $M$ (see supplement for the proposition and proof). In practice, we impose three constraints (see Section~\ref{sec:method}): (i) bound $M$ to $[M_{\min},M_{\max}]=[44/3,44]$; (ii) require $\Theta$ to be positive and monotone; and (iii) use a low-degree local-polynomial representation of $M$.
The simulation study in Section~\ref{sec:simulation} provides empirical evidence that these constraints can help recover the true latent surfaces.

\subsection{Direct Interpolation of EOS Table}
\label{sec:direct_interpolation}
An intuitive and alternative approach to ours is to omit the free energy altogether and instead model $E_\mathrm{ion}$ and $P_\mathrm{ion}$ directly by smoothing the observed values using semi-parametric smoothing methods \citep{wahba1990spline,ruppert2003semiparametric,wood2017generalized,ramsay1988monotone}, with each surface represented as a regularized basis expansion. Concretely, we implement this direct approach with a generalized additive model (GAM) using tensor-product splines and a log link function, to capture the joint temperature--density dependence and to preserve non-negativity of the outputs, respectively. However, this kind of default implementation cannot guarantee thermodynamic consistency, because $E_{\mathrm{ion}}$ and $P_{\mathrm{ion}}$ are fit as two separate surfaces rather than as derivatives of a shared free energy. As a result, the fitted surfaces are not guaranteed to satisfy the thermodynamic derivative relations, such as those given in \citet{myint2017free}, which ultimately arise from the equality of mixed partial derivatives (Clairaut's theorem).

\begin{figure}[!htb]
    \centering
    \includegraphics[width=\textwidth]{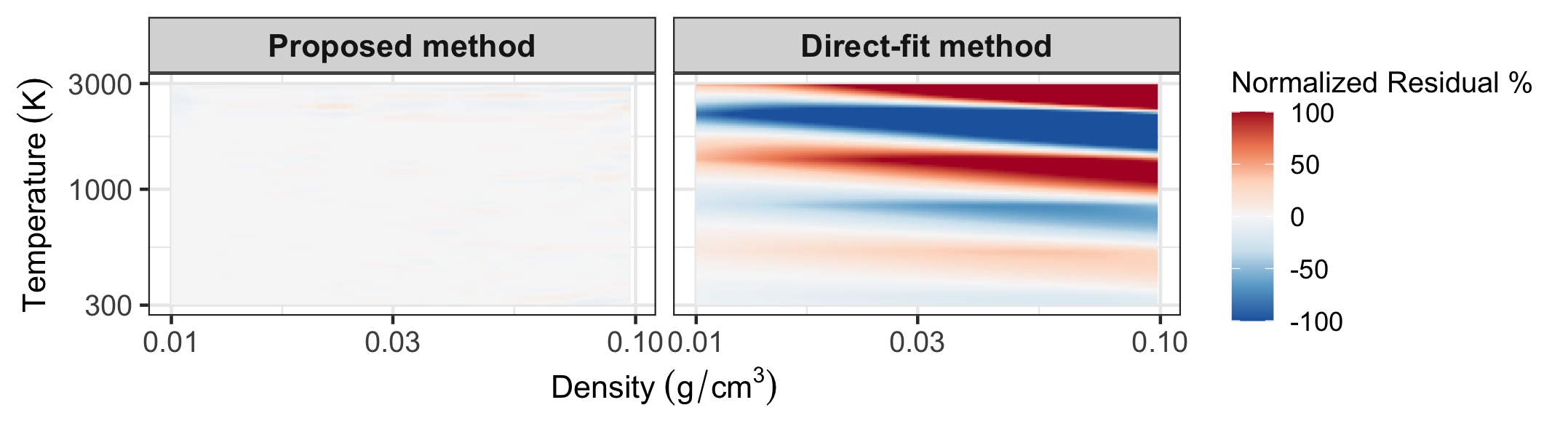}
    \caption{Consistency check over $\rho \in [0.01, 0.1]~\mathrm{g/cm^3}$ and $T \in [300, 3000]~\mathrm{K}$ for our proposed free-energy-based approach (left) and direct GAM smoothing of $E_{\mathrm{ion}}$ and $P_{\mathrm{ion}}$ (right). Color shows $100\,[T(\partial^2 P/\partial T^2)_V - (\partial C_V/\partial V)_T]/S$, where $S$ is the median of $|T(\partial^2 P/\partial T^2)_V|$ and $|(\partial C_V/\partial V)_T|$, pooled across both methods so the two panels share a scale.}
    \label{fig:direct_approach_inconsistency}
\end{figure}

We illustrate this inconsistency using Equation~(4) of \citet{myint2017free}, $(\partial C_V/\partial V)_T = T(\partial^2 P/\partial T^2)_V$, where $C_V = (\partial E/\partial T)_V$. This identity must hold if $P_{\mathrm{ion}}$ and $E_{\mathrm{ion}}$ are derived from a common free energy, since it follows from the equality of mixed partial derivatives. We compare this direct GAM fit against our proposed approach (described in Section~\ref{sec:method}), which instead derives $E_{\mathrm{ion}}$ and $P_{\mathrm{ion}}$ jointly as derivatives of a single, shared free-energy function, so that this consistency relation holds automatically rather than needing to be checked. For each approach, we numerically differentiate the fitted surfaces to evaluate $(\partial C_V/\partial V)_T$ and $T(\partial^2 P/\partial T^2)_V$ and compare them across the domain. Figure~\ref{fig:direct_approach_inconsistency} shows that our proposed method's residual is negligible throughout the entire domain, while the direct-fit GAM surface shows systematic banded violations of tens to 100\% of the shared scale $S$: independently smoothing $E_{\mathrm{ion}}$ and $P_{\mathrm{ion}}$ does not enforce thermodynamic consistency and is therefore not physically valid.

\section{Partitioned EOS Estimation}
\label{sec:method}
We now introduce a partitioned procedure for fitting the observation model~\eqref{eq:obs_model} by jointly estimating the global Debye temperature curve $\Theta(x)$ and local molar mass surface $M(x,y)$. We first describe the individual stages and then combine them into the full EOS estimation framework.

\subsection{Stages for Partitioned EOS Estimation}

\paragraph{(i) Partitioning of EOS Tables}
Partition-based, or divide-and-conquer, modeling decomposes a complex global model into simpler localized ones that are estimated separately and recombined, often with substantial computational savings \citep{Zhang2015,chen2021divide}. Beyond scalability, partitioning accounts for nonstationarity by allowing local behavior to vary in smoothness and heterogeneity across partitions, as opposed to global stationary models. Common instantiations include partitioned Gaussian process models such as treed Gaussian processes \citep{gramacy2008bayesian}, piecewise Gaussian processes \citep{KimMallickHolmes2005}, and locally nonstationary basis expansions \citep{lee2023scalable}, where each region is locally parameterized, while the aggregate surface remains nonstationary.

In the log-domain $(x,y)$, we divide the table into $H=\left\lfloor N/n_p \right\rfloor$ rectangular regions, where $N$ is the total number of rows in the EOS table (i.e., the number of distinct temperature points at which the table is tabulated) and $n_p$ is the desired number of records in each partition. Each partition $\mathcal{K}_h$, for $h = 1,\dots,H$, is defined by bounds, $(x_{l,h},x_{u,h})$ and $(y_{l,h},y_{u,h})$, and buffer $\delta > 0$:
\[
\mathcal{K}_h :=
\left\{i:
x_{l,h} - \delta \le x_i \le x_{u,h} + \delta,
\quad
y_{l,h} - \delta \le y_i \le y_{u,h} + \delta
\right\}, \qquad h = 1,\dots,H.
\]
Adjacent partitions overlap in a buffer zone of width $\delta$, so a data point near a boundary may belong to more than one partition. Partitioning is computed once at the outset.

\paragraph{(ii) Localized Molar Mass Surfaces}
Within each partition $\mathcal{K}_h$, we model the partition-specific predictor $\eta_{M,h}(x,y)$ as a polynomial basis expansion in the partition-centered coordinates $(x - x_h, y - y_h)$, where $(x_h,y_h)$ denote the partition centers. We write 
\[ \eta_{M,h}(x,y) = \mathbf{b}_{d}(x-x_h,y-y_h)^\top\mathbf{a}_h, \] 
where $\mathbf{b}_{d}$ contains the polynomial basis terms for degree $d$ and $\mathbf{a}_h$ is the corresponding coefficient vector. We consider bilinear, quadratic, and cubic specifications with 4, 6, and 10 coefficients, respectively.

Physical bounds are enforced through a logistic transformation: \[ M_h(x,y) = M_{\min} + (M_{\max}-M_{\min})\,\sigma\!\big(\eta_{M,h}(x,y)\big), \qquad \sigma(u)=\frac{1}{1+e^{-u}}, \] where $M_{\min}=44/3$ and $M_{\max}=44$ are the physically motivated lower and upper bounds on the molar mass introduced in Section~2. 

\paragraph{(iii) Debye Temperature Estimation using I-Splines}
We represent the Debye temperature as a positive, monotone function of density using the I-spline expansion in $x$:
\[
\eta_\Theta(x)=\beta_0+\sum_{j=1}^{J}\beta_j\, I_j(x),
\qquad
\Theta(x)=\exp\!\big(\eta_\Theta(x)\big),
\qquad \beta_j\ge 0, \ \ j\ge 1,
\]
where $\{I_j\}_{j=1}^{J}$ are cubic I-spline basis functions with interior knots placed at quantiles of the observed densities and $\boldsymbol{\beta}=(\beta_0,\dots,\beta_J)$ are global coefficients shared across all partitions.

I-spline bases ensure monotonicity by construction \citep{ramsay1988monotone} and are defined as
$I_j(x)=\int_{x_{\min}}^{x} M_j(u)\,\mathrm{d}u$ where M-splines $\{M_j\}$ are a nonnegative, unit-integral rescaling of the B-spline basis on a given knot sequence \citep{curry1988polya}. Each $I_j$ is monotone nondecreasing and ranges from $0$ to $1$ over the knot span since each $M_j$ integrates to one. Constraining $\beta_j \ge 0$ ensures $\Theta(x)$ is monotone in density. Shape-restricted parameterizations are common in monotone curve estimation \citep{ramsay1988monotone,meyer2008inference,xu2023bayesian}. In this case, the intercept $\beta_0$ sets the overall level, the exponential map enforces positivity, and knot placement controls curve flexibility. Since M-spline bases are smooth, the density derivative $d\Theta/dx$ required by thermodynamic properties is available in closed form.


\subsection{EOS Framework for Inferring Molar Mass and Debye Temperature}
\label{sec:coupled}

We now synthesize the previous three components into a single estimation procedure that fits the global Debye coefficients $\boldsymbol{\beta}$ and the partition-wise molar mass coefficients $\{a_{j,h}\}$ through a single loss function coupling $E_{\mathrm{ion}}$ and $P_{\mathrm{ion}}$. Since $\Theta$ is shared across the full table while each $M_h$ is local, we alternate between global Debye updates and parallel partition-wise molar mass updates until convergence. The full molar mass surface $M(x_i,y_i)$ is obtained by aggregating the partition-specific models with the overlap weights defined below.

\paragraph{Loss function.}
\label{par:loss}
We fit the model by minimizing the mean squared relative error of the ion-thermal internal energy and pressure observables. Let $(E_{\mathrm{ion},i},P_{\mathrm{ion},i})$ denote the tabulated values at grid point $i$, and let $(E^{\ast}_{\mathrm{ion},i},P^{\ast}_{\mathrm{ion},i})$ denote the corresponding fitted values obtained from the current $(\Theta,M)$ through the Cowan free energy \eqref{eq:Cowan} and the thermodynamic identities \eqref{eq:InternalEnergy}--\eqref{eq:Pressure}. The loss is
\begin{equation}
\label{eq:loss}
\mathcal{L}(\Theta,M)
=
\frac{1}{N}\sum_{i=1}^N
\left[
\left(
\frac{E^{\ast}_{\mathrm{ion},i}-E_{\mathrm{ion},i}}{\max\!\big(\varepsilon,|E_{\mathrm{ion},i}|\big)}
\right)^{\!2}
+
\left(
\frac{P^{\ast}_{\mathrm{ion},i}-P_{\mathrm{ion},i}}{\max\!\big(\varepsilon,|P_{\mathrm{ion},i}|\big)}
\right)^{\!2}
\right],
\end{equation}
where $\varepsilon>0$ is a small constant that prevents division by zero and reduces sensitivity where $|E_{\mathrm{ion},i}|$ or $|P_{\mathrm{ion},i}|$ is near zero. Minimizing \eqref{eq:loss} is maximum likelihood estimation under the observation model \eqref{eq:obs_model} with independent Gaussian errors $\varepsilon^{E}_i$ and $\varepsilon^{P}_i$ that have heteroskedastic standard deviations $\nu\max(\varepsilon, |E_{\mathrm{ion},i}|)$ and $\nu\max(\varepsilon, |P_{\mathrm{ion},i}|)$ with a common relative-noise scale $\nu$ and floor $\varepsilon$. 

\paragraph{Overlapping Partition Aggregation.}
Adjacent partitions are deliberately overlapped to avoid discontinuities at the boundaries. Since partitions are fit independently, non-overlapped boundaries can produce abrupt changes in the predictor. Overlapping blends predictions from neighboring partitions, yielding smoother transitions across boundaries.

A row $i$ may belong to multiple partitions, indexed by
$\mathcal{H}_i=\{h:i\in\mathcal{K}_h\}$. Each partition $h$ contributes a latent predictor
$\eta_{ih}:=\eta_{M,h}(x_i,y_i)$ and its derivatives. Predictions are combined using weights
$w_{ih}=k_{ih}q_h$, where $k_{ih}=\exp(-\tfrac12 d_{ih}^2)$, with
$d_{ih}^2=\left(\frac{x_i-c_{x,h}}{s_{x,h}}\right)^2+
\left(\frac{y_i-c_{y,h}}{s_{y,h}}\right)^2$, favors points near the partition center, and
$q_h=(\varepsilon_q+\mathcal{L}_h)^{-1}$ favors partitions with smaller loss
\eqref{eq:loss}. The aggregated latent predictor is $\bar\eta_i = \sum_{h\in\mathcal{H}_i} w_{ih}\,\eta_{ih} / \sum_{h\in\mathcal{H}_i} w_{ih}$, and its density and temperature derivatives are aggregated with the same weights, $\partial_\bullet\bar\eta_i = \sum_{h\in\mathcal{H}_i} w_{ih}\,\partial_\bullet\eta_{ih} / \sum_{h\in\mathcal{H}_i} w_{ih}$ for $\bullet\in\{x,y\}$. Here, the weights are treated as locally constant when differentiating. Finally,
$\bar\eta_i$ is mapped to molar mass through
$M(x_i,y_i)=M_{\min}+(M_{\max}-M_{\min})\sigma(\bar\eta_i)$,
with derivatives obtained by the chain rule.

\paragraph{Estimation algorithm.}
Estimation proceeds by alternating between a global update of the Debye curve and a partition-wise update of the molar mass surface (see Algorithm~\ref{alg:outer_inner}). The molar mass surface is initialized at the midpoint $M^{(0)}(x,y)=(M_{\min}+M_{\max})/2$. At iteration $k$, the Debye curve $\Theta^{(k)}(x)$ is re-estimated as a monotone spline by minimizing the global relative error loss \eqref{eq:loss} given $M^{(k-1)}$. Then, the molar mass surface is updated independently within each overlapping partition and recombined into a candidate $M^{(k)}_{\mathrm{cand}}$. The candidate is accepted only if it does not increase the full-table loss, and is otherwise damped toward $M^{(k-1)}$ by backtracking on a convex-combination weight $\omega\in(0,1]$ (Algorithm~\ref{alg:outer_inner}). The resulting loss sequence is non-increasing and bounded below. The algorithm stops once changes in the: (i) loss function; (ii) Debye I-spline coefficients; and (iii) molar mass surface remain below set tolerances for a pre-specified number of consecutive iterations, set to three in this study.

\begin{algorithm}[htb]
\caption{Alternating (block-coordinate) estimation of $(\Theta,M)$}
\label{alg:outer_inner}
\KwIn{EOS table $\{(E_{\mathrm{ion},i},P_{\mathrm{ion},i})\}_{i=1}^N$; partitions $\{\mathcal{K}_h\}_{h=1}^H$; bounds $[M_{\min},M_{\max}]$; step controls $\omega_{\mathrm{start}},\omega_{\mathrm{shrink}},\omega_{\min}$; convergence controls \texttt{min\_outer}, \texttt{patience}, tolerances}
\KwOut{Estimated Debye coefficients $\hat{\boldsymbol{\beta}}$ (curve $\hat{\Theta}$) and molar mass coefficients $\{\hat{a}_{j,h}\}$ (surface $\hat{M}$)}

\SetKwFunction{Optim}{L-BFGS-B}

Initialize $M^{(0)}\!\leftarrow\!(M_{\min}+M_{\max})/2$; set $k\leftarrow 0$\;

\Repeat{convergence criteria hold for \texttt{patience} consecutive iterations \textbf{and} $k\ge\texttt{min\_outer}$}{
    $k \leftarrow k+1$\;

    \textbf{Global Debye update:} $\ \boldsymbol{\beta}^{(k)} \leftarrow \arg\min_{\boldsymbol{\beta}}\ \mathcal{L}\!\left(\Theta(\cdot;\boldsymbol{\beta}),\,M^{(k-1)}\right)$ subject to $\beta_j\ge 0\ (j\ge1)$, giving $\Theta^{(k)}$\;

    \textbf{Partition-wise $M$ update (parallel):} \For{$h=1,\dots,H$ \textnormal{\textbf{in parallel}}}{
        $\{a_{j,h}\}^{(k)} \leftarrow$ \Optim{minimize $\mathcal{L}$ restricted to rows $i\in\mathcal{K}_h$, with $\Theta^{(k)}$ fixed}\;
    }
    Recombine into candidate surface $M^{(k)}_{\mathrm{cand}}$\;

    \textbf{Monotone acceptance (backtracking):} $\ \omega\leftarrow\omega_{\mathrm{start}}$\;
    \While{$\omega\ge\omega_{\min}$}{
        $M^{(k)}_{\mathrm{try}}\leftarrow(1-\omega)M^{(k-1)}+\omega M^{(k)}_{\mathrm{cand}}$\;
        \lIf{$\mathcal{L}(\Theta^{(k)},M^{(k)}_{\mathrm{try}})\le \mathcal{L}(\Theta^{(k-1)},M^{(k-1)})$}{$M^{(k)}\leftarrow M^{(k)}_{\mathrm{try}}$; \textbf{break}}
        $\omega\leftarrow\omega\cdot\omega_{\mathrm{shrink}}$\;
    }
    \lIf{no $\omega$ accepted}{$M^{(k)}\leftarrow M^{(k-1)}$ \textnormal{(reject)}}

    Record $\mathcal{L}^{(k)}=\mathcal{L}(\Theta^{(k)},M^{(k)})$ and parameter-update sizes\;
}

\Return{$\hat{\boldsymbol{\beta}}=\boldsymbol{\beta}^{(k)},\ \{\hat{a}_{j,h}\}=\{a_{j,h}\}^{(k)}$}\;
\end{algorithm}

\paragraph{Prediction at unobserved states.}
The fitted model predicts $E_{\mathrm{ion}}$ and $P_{\mathrm{ion}}$ at any unobserved state $(x^{*},y^{*})$ in the EOS table. Each partition covering $(x^{*},y^{*})$ yields a local predictor $\eta_{M,h}^{*}$, and these are combined by the same overlap weights $w_{ih}$ used during estimation, giving an aggregate $\bar\eta^{*}$ and a molar mass $M(x^{*},y^{*})=M_{\min}+(M_{\max}-M_{\min})\,\sigma(\bar\eta^{*})$. Evaluating the Cowan free energy at $M(x^{*},y^{*})$ and the global Debye curve $\Theta(x)$ yields the predicted $E_{\mathrm{ion}}^{*}$ and $P_{\mathrm{ion}}^{*}$ through the thermodynamic identities in \eqref{eq:InternalEnergy} and \eqref{eq:Pressure}.

\subsection{Implementation Details}\label{subsec:impl}


We model $\Theta(x)$ with cubic I-splines and its derivative with the corresponding M-splines, using 20 equally spaced interior knots and boundary knots at the endpoints of the $x$ domain. The bases are constructed using functions in the \texttt{splines2} R package \citep{wang2021shape}. For the partitioned estimation of the molar mass surface, the partition-level fits are distributed across 20 processors. All computations were executed on the Hopper cluster at George Mason University, using nodes each equipped with a single 2.4 GHz Intel Xeon Gold 6240R processor. 

To assess sensitivity to the partitioning scheme, we vary three settings: the target partition size $n_p$; the overlap width $\delta$; and the polynomial degree of the localized approximations $\eta_{M,h}(x,y)$. These respectively control the number of partitions, the amount of overlap, and the smoothness of the local molar mass surface. We consider $n_p\in\{10,15,20,25,50,100\}$, $\delta\in\{0.01,0.05,0.1\}$, and bi-linear, quadratic, and cubic degrees. In Sections~\ref{sec:simulation} and \ref{sec:casestudy} we present the results of the sensitivity analysis and adopt the combination with the lowest relative-error loss from this sensitivity analysis.

We cap the alternating procedure at $500$ outer iterations and each L-BFGS-B optimization at $1000$ iterations. For the damped $M(x,y)$-surface update, backtracking begins at $\omega=1$ and multiplies $\omega$ by $0.7$ until the full-table loss decreases or $\omega<1/16$. Convergence is declared once, after a minimum of three outer iterations, the following conditions are satisfied for two consecutive iterations: (i) the relative change in the objective function is less than $10^{-3}$; (ii) the maximum absolute change (i.e., infinity norm) in the global spline coefficients is less than $10^{-3}$; and (iii) the relative root-mean-square change in $M$ is less than $2\times 10^{-3}$. To compute the partition score $q_h$, we set $\varepsilon_q = 10^{-8}$.

\section{Uncertainty Quantification}\label{sec:UQ}

The estimation procedure of Section~\ref{sec:method} yields point estimates of $M(x,y)$, $\Theta(x)$, and the derived $E_{\mathrm{ion}}$ and $P_{\mathrm{ion}}$. We quantify the underlying uncertainty with a parametric bootstrap that resamples from the fitted model under a heteroskedastic relative-error structure matching the loss~\eqref{eq:loss}. 

\subsection{Parametric bootstrap}\label{sec:bootstrap}
Starting from a baseline fit (Algorithm~\ref{alg:outer_inner}), we estimate the local relative-error scale over the $(x,y)$ domain. For each observation, the relative residuals are
  \[
  r_{E,i} = \frac{E_{\mathrm{ion},i} - \hat{E}_{\mathrm{ion},i}}{s_{E,i}},
  \qquad
  r_{P,i} = \frac{P_{\mathrm{ion},i} - \hat{P}_{\mathrm{ion},i}}{s_{P,i}},
  \]
  with scales $s_{E,i} = \max(c_E, |\hat{E}_{\mathrm{ion},i}|)$ and $s_{P,i} = \max(c_P, |\hat{P}_{\mathrm{ion},i}|)$, where $c_E,c_P$ play the same role as $\varepsilon$ in \eqref{eq:loss}. Smoothing the squared relative residuals over $(x,y)$ gives location-dependent relative standard deviations $\hat{\sigma}_E(x,y)$ and $\hat{\sigma}_P(x,y)$.

For each replicate $b = 1,\ldots,B$, we generate synthetic responses
  \[
  E_{\mathrm{ion},i}^{(b)} = \hat{E}_{\mathrm{ion},i} + s_{E,i}\hat{\sigma}_{E}(x_i,y_i) Z_{E,i}^{(b)},
  \qquad
  P_{\mathrm{ion},i}^{(b)} = \hat{P}_{\mathrm{ion},i} + s_{P,i}\hat{\sigma}_{P}(x_i,y_i) Z_{P,i}^{(b)},
  \]
  with $Z_{E,i}^{(b)}, Z_{P,i}^{(b)} \stackrel{\mathrm{iid}}{\sim} N(0,1)$,
  preserving the fitted mean while letting the error variance vary across the domain. Each replicate is refit with the same alternating procedure (Section~\ref{sec:method}), with the partitioning scheme, polynomial degree, and tolerances held fixed, initialized at $\hat{M}$ and $\hat{\Theta}$, and run until the convergence criteria of Algorithm~\ref{alg:outer_inner} are met. 
  
  The $B$ refits yield ensembles of $M$, $\Theta(x)$, $E_{\mathrm{ion}}$, and $P_{\mathrm{ion}}$, and pointwise $95\%$ bootstrap intervals are given by the $2.5$th and $97.5$th percentiles of each ensemble. The percentile method makes no normality assumption and accommodates the asymmetric sampling distributions induced by the bounded molar mass surface and monotone Debye-temperature constraints. Details regarding prediction bands, residual-scale smoothing, and numerical stabilization are described in the Supplement.

\subsection{Parametric Bootstrap Implementation Details}

In our implementation, bootstrap generation uses a fixed random seed for reproducibility, and each replicate is refit independently using parallel computation across overlapping partitions. We use $B=100$ bootstrap replicates throughout this study. Increasing $B$ can improve the stability of the estimated bootstrap intervals \citep{efron1994introduction}, though at greater computational cost.
The squared relative residuals are smoothed on the log scale, which guarantees positive variance estimates. For energy and pressure alike, we fit a bivariate local quadratic regression (\texttt{loess} in R, degree $2$, span $0.10$) of $\log(r_{E,i}^2 + \epsilon_E)$ and $\log(r_{P,i}^2 + \epsilon_P)$ on $(x,y)$. The small offsets $\epsilon_E$ and $\epsilon_P$ are the first percentiles of the strictly positive $r_{E,i}^2$ and $r_{P,i}^2$, respectively, and prevent $\log 0$. Let $\ell_E(x,y) = \log\!\big(r_{E}^2(x,y)+\epsilon_E\big)$ and $\ell_P(x,y) = \log\!\big(r_{P}^2(x,y)+\epsilon_P\big)$. The fitted standard deviations are
$\hat{\sigma}_E(x,y) = \sqrt{\exp\{\hat{\ell}_E(x,y)\}}$ and $\hat{\sigma}_P(x,y) = \sqrt{\exp\{\hat{\ell}_P(x,y)\}}$ each with a floor of $10^{-6}$ for numerical stability. The polynomial degree and span may be adjusted to control the smoothness of the estimated surfaces. The lower empirical quantile levels used to determine $c_E$ and $c_P$ are both set to $0.01$. These levels may be adjusted on a case-by-case basis.

\section{Simulation Study}
\label{sec:simulation}
Before we proceed to the CO$_2$ test case in Section~\ref{sec:casestudy}, we demonstrate our proposed partitioned approach on a simulation study involving multiple synthetic EOS tables with known latent surfaces to assess how well $E_{\mathrm{ion}}$ and $P_{\mathrm{ion}}$ are recovered. We compare it primarily against the Global Cowan--radial basis function (RBF) model, a nonpartitioned model that isolates the effect of replacing a global representation of $M(x,y)$ with localized polynomial representations and overlap aggregation. We vary the partitioning scheme to examine the tradeoff between flexibility and global consistency and quantify the estimation uncertainty for each case.
\subsection{Design}
\label{sec:design}
We generated 12 synthetic molar mass surfaces from baseline forms $g(x,y)$ representing distinct types of spatial dependence: monotone trends in $x$ and/or $y$, mixed trends of opposing sign, parabolic curvature, and sinusoidal variation. Each baseline surface is rescaled onto $[M_{\min},M_{\max}]$
\begin{equation*}
M(x,y) = M_{\min} + (M_{\max}-M_{\min})\,
\frac{g(x,y)-\min_{x,y} g}{\max_{x,y} g - \min_{x,y} g},
\label{eq:sim_M_rescale}
\end{equation*}
so that, by construction, the resulting surface depends only on the case type and not on the generating constants; we use $s=1$ and $b=0$ throughout. The twelve resulting molar mass surfaces are shown in Figure~\ref{fig:molarmass}, and the full list of baseline forms is given in the Supplement (Table~S4).

The Debye temperature $\Theta(x)$, shared across all cases, is built from a cubic I-spline expansion in $x$ with nonnegative coefficients ($\beta_j\ge 0$, $j\ge1$, drawn once at random) so that $\eta_\Theta(x)=\beta_0+\sum_{j=1}^J \beta_j I_j(x)$ is monotone nondecreasing; exponentiating and rescaling gives
\begin{equation*}
\Theta(x) = \Theta_{\min} + (\Theta_{\max}-\Theta_{\min})\,
\frac{e^{\eta_\Theta(x)}-\min_{x} e^{\eta_\Theta}}{\max_{x} e^{\eta_\Theta}-\min_{x} e^{\eta_\Theta}},
\end{equation*}
with $(\Theta_{\min},\Theta_{\max})=(300,5000)$~K and knots at density quantiles. $\Theta(x)$ is drawn once under a fixed seed and shared across all tables, so differences across cases reflect only the molar mass surface and the noise; the resulting curve is shown in Figure~\ref{fig:debye_sim}.

All surfaces are evaluated on the same grid as the CO$_2$ case study ($N_\rho=100$ densities $\times$ $N_T=200$ temperatures, $N=20{,}000$ points), spanning $x\in[-1.95,-0.05]$ and $y\in[2.35,4.13]$. The simulated $M(x,y)$ and $\Theta(x)$ are fed into the Cowan free energy \eqref{eq:Cowan} ($\alpha = 2$, $R = 8.3145\times10^{-5}$ Mbar$\cdot$cm$^3$$/$mol$/$K), which returns $E_{\mathrm{ion}}$, $P_{\mathrm{ion}}$, and $S_{\mathrm{ion}}$ in closed form. We perturb $E_{\mathrm{ion}}$ and $P_{\mathrm{ion}}$ at relative noise level $\nu\in\{0,0.01\}$ (noiseless and $1\%$ noise); full details of the noise model are given in the Supplement. Crossing the two noise levels with the twelve surfaces produces $12\times2=24$ EOS tables.

\begin{figure}[!htb]
\centering
\begin{subfigure}{0.4\textwidth}
\centering
\includegraphics[width=\textwidth]{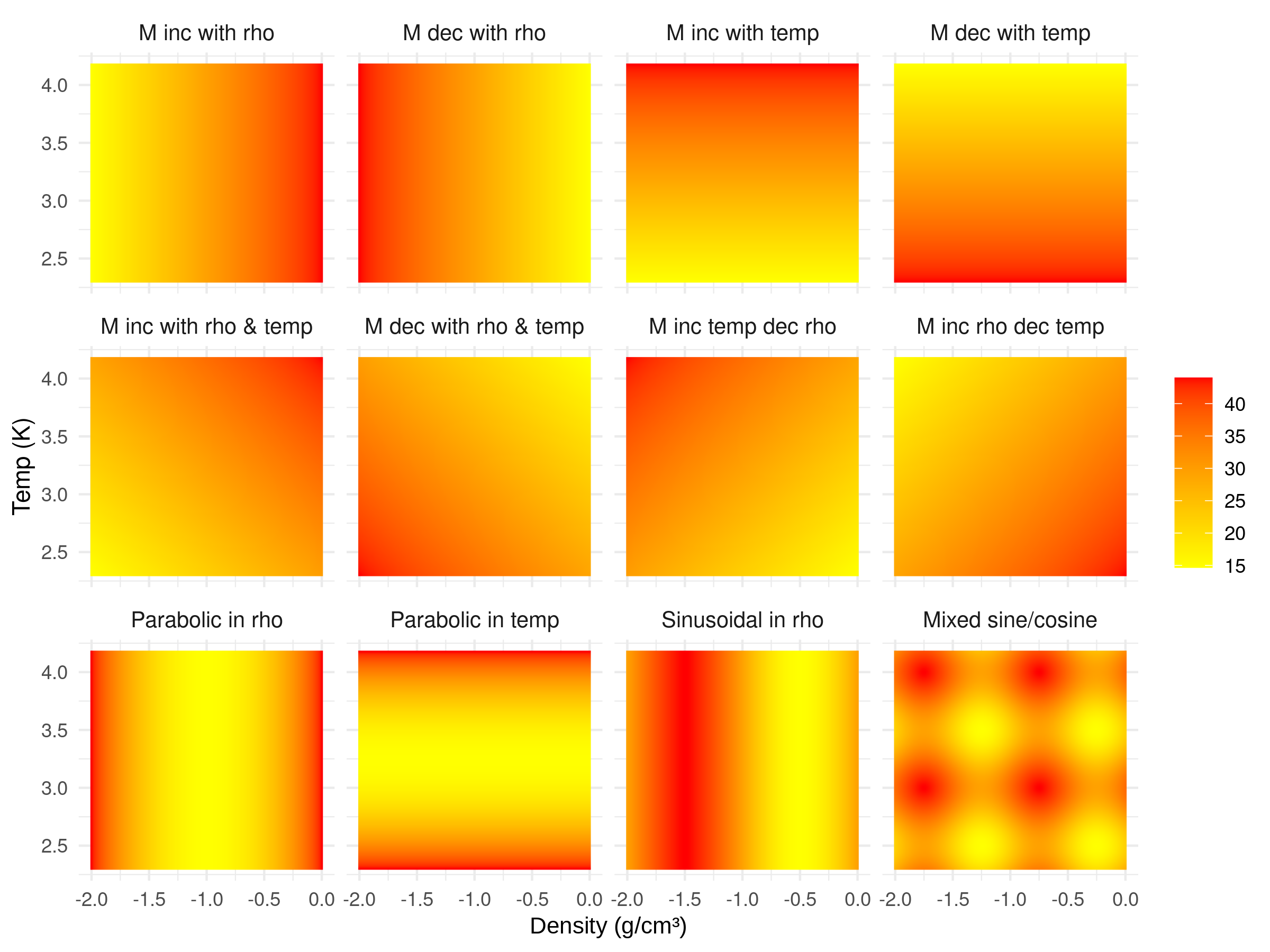}
\caption{Synthetic molar mass surfaces, cases 1--12.}
\label{fig:molarmass}
\end{subfigure}
\hfill
\begin{subfigure}{0.4\textwidth}
\centering
\includegraphics[width=\textwidth]{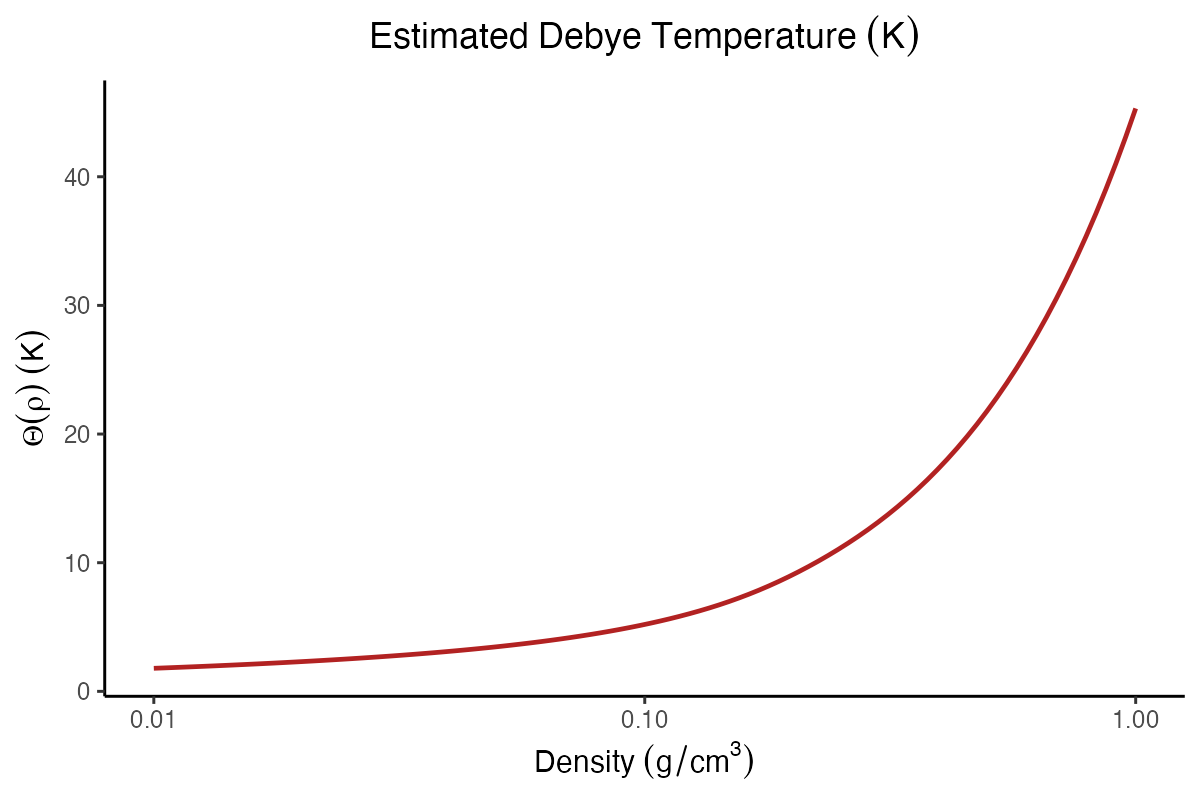}
\caption{Debye temperature $\Theta(\rho)$, shared across cases.}
\label{fig:debye_sim}
\end{subfigure}
\caption{Simulation design components: (a) the twelve baseline molar mass surfaces and (b) the shared Debye temperature curve fit via natural cubic spline.}
\label{fig:design_panels}
\end{figure}

The partitioned approach fits polynomial functions (bilinear, quadratic, cubic) on overlapping partitions, while the Global Cowan--RBF model uses a single Gaussian RBF expansion over the full domain (see Supplement). We reserve $\sim15\%$ of grid locations as a regularly spaced test subgrid within the central $95\%$ of the $x$- and $y$-ranges, with the remainder used for training. Prediction uncertainty is quantified via the parametric bootstrap of Section~\ref{sec:UQ} using $B=100$ replicates. We report relative errors for $E_{\mathrm{ion}}$ and $P_{\mathrm{ion}}$ on both sets, the prediction interval coverage probability (PICP) of the nominal $95\%$ prediction intervals \citep{khosravi2010lower, pearce2018high}, and the interval score \citep{gneiting2007strictly}.

\subsection{Results}
We first conducted a sensitivity analysis for a single model fit across partition sizes $n_p \in \{10,\allowbreak 15,\allowbreak 20,\allowbreak 25,\allowbreak 50,\allowbreak 100\}$, overlap fractions $\delta \in \{0.01,\allowbreak 0.05,\allowbreak 0.10\}$, polynomial bases (linear, quadratic, and cubic), and relative noise standard deviations $\nu \in \{0,\allowbreak 0.01\}$. Smaller partition sizes and lower overlap fractions, particularly $\delta = 0.01$, generally produced lower relative error losses for held-out data, and the cubic basis performed well across most settings compared to quadratic and linear. Complete results are provided in the Supplementary Material.
\input{bootstrap_band_scores_all.tex}

\begin{figure}[!htb]
  \centering
  \includegraphics[width=\textwidth]{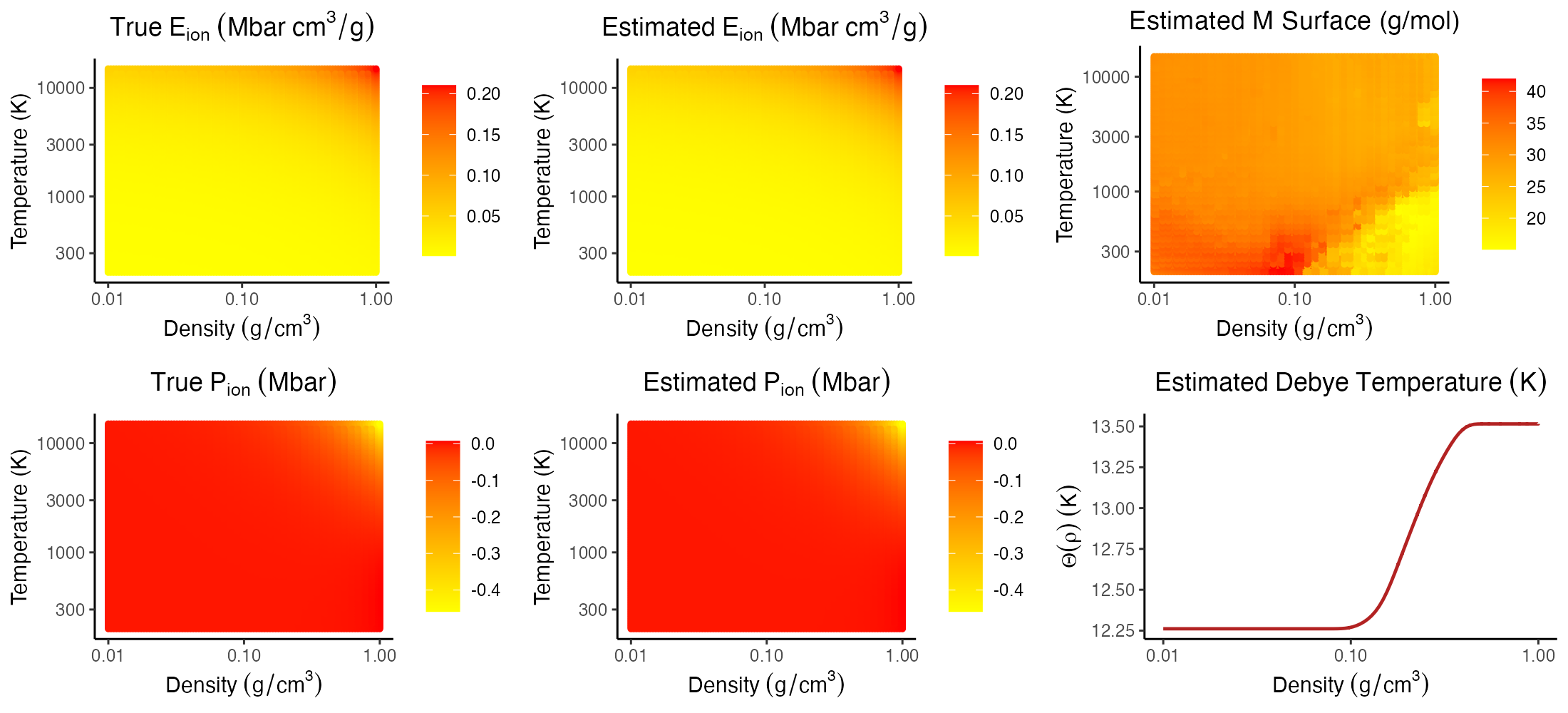}
  \caption{True and estimated thermodynamic surfaces and estimated latent quantities for Case~2 of the simulation study using cubic local polynomials with $\nu=0$, $n_{\mathrm{pp}}=15$, and $\delta=0.01$. The top row shows the true and estimated $E_{\mathrm{ion}}$ (Mbar\,cm$^{3}$/g) and $M$ (g/mol). The bottom row shows the true and estimated $P_{\mathrm{ion}}$ (Mbar) and  $\Theta(\rho)$ (K).}
  \label{fig:sim_case3_surfaces}
\end{figure}

\begin{figure}[!htb]
\centering
\includegraphics[width=\textwidth]{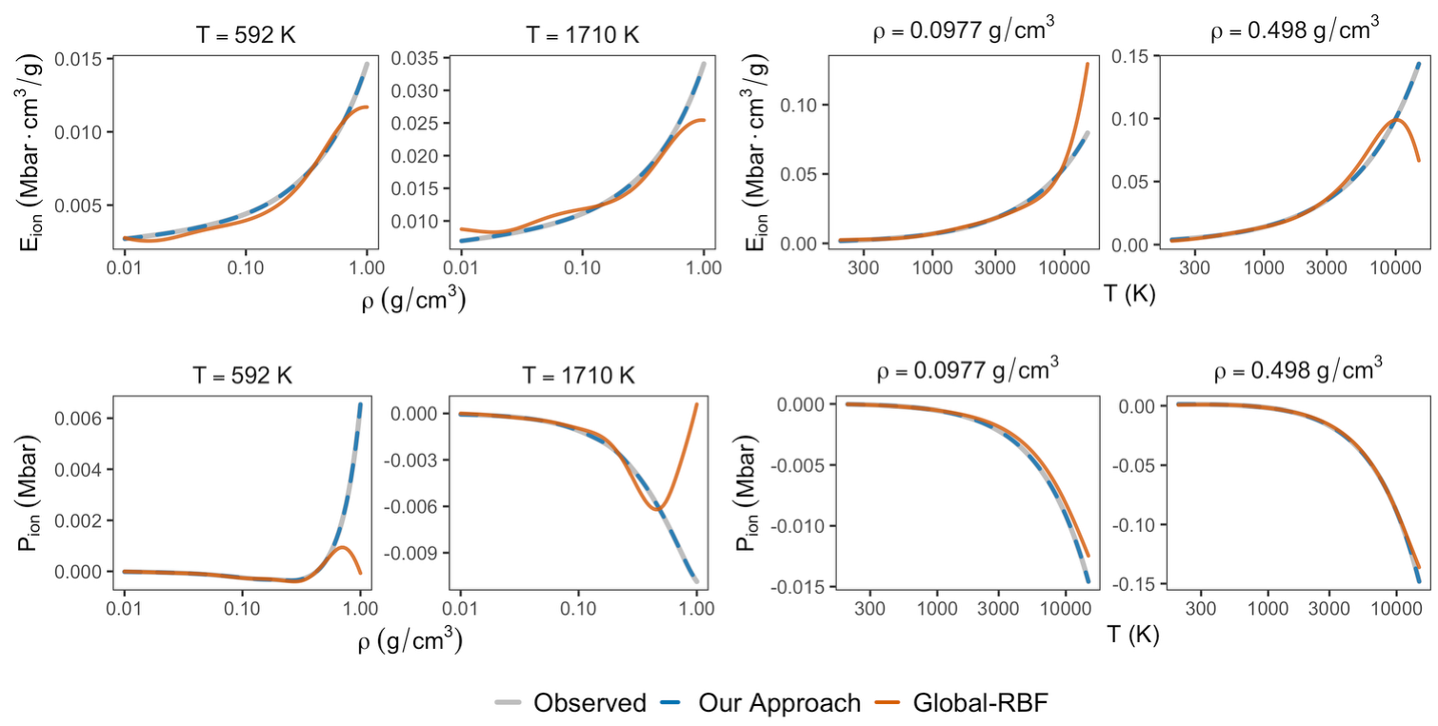}
\caption{Simulation study, Case 2 ($M$ decreases with $x$): comparison of observed and predicted ion-thermal energy $E_{\mathrm{ion}}$ (top row) and pressure $P_{\mathrm{ion}}$ (bottom row) from our partitioned approach and the Global Cowan--RBF model. The left two columns show isotherms at $T=592$~K and $T=1710$~K, plotted as a function of density $\rho$; the right two columns show isochores at $\rho=0.0977$~g/cm$^3$ and $\rho=0.498$~g/cm$^3$, plotted as a function of temperature $T$. Our approach (blue dashed) closely tracks the observed values (gray) along both isotherms and isochores, while the Global-RBF model (orange) shows visible deviations, particularly in $E_{\mathrm{ion}}$ at low density and in the high-density/high-temperature tails of $P_{\mathrm{ion}}$.}
\label{fig:compare_case2_methods}
\end{figure}
Based on test relative error loss and visual inspection of the fitted $E_{\mathrm{ion}}$ and $P_{\mathrm{ion}}$ surfaces, we selected the model settings used for uncertainty quantification. Across nearly all cases and noise levels, the partitioned approach yielded lower relative error loss than the Global Cowan--RBF model (Table~\ref{tab:band_scores_all}). Figure~\ref{fig:sim_case3_surfaces} compares the true and estimated surfaces for Case~2, and Figure~\ref{fig:compare_case2_methods} compares both methods along representative isotherms and isochores. Our approach closely tracks the observed values, while the Global Cowan--RBF model deviates visibly, particularly in $E_{\mathrm{ion}}$ at low density and in the high-density, high-temperature tails of $P_{\mathrm{ion}}$. This gap arises because a single Gaussian RBF expansion imposes one length scale across the domain, whereas the molar-mass surface varies at multiple scales; as discussed in Section~\ref{sec:intro}, each local partition instead adapts to the smoothness of its own region.

\begin{figure}[!htb]
    \centering
    \includegraphics[width=0.9\textwidth]{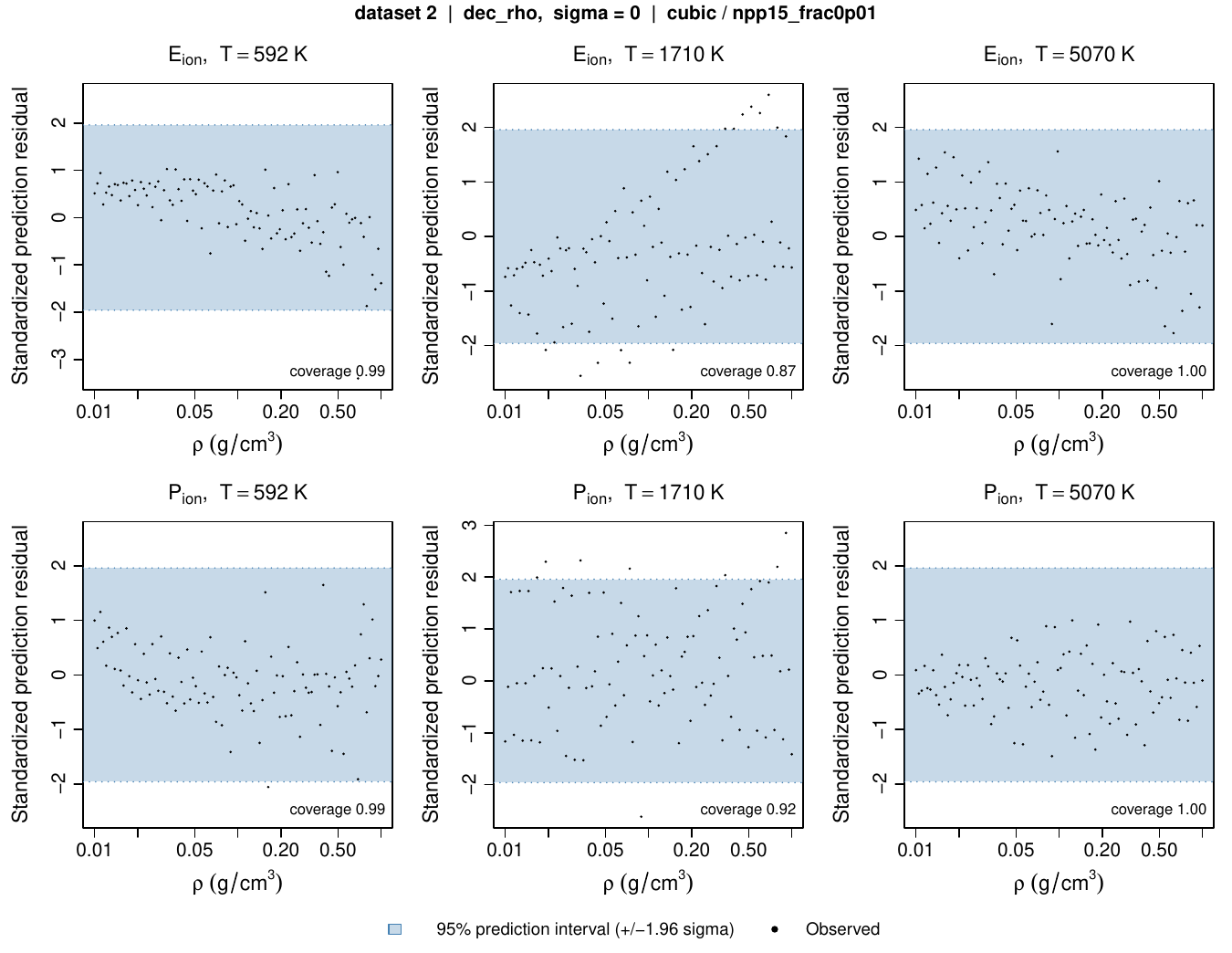}
    \caption{Standardized prediction residuals for $E_{\mathrm{ion}}$ (top row) and $P_{\mathrm{ion}}$ (bottom row) as a function of density $\rho$, for simulation study case 2, shown along three representative isotherms ($T = 592$, $1710$, and $5070$~K). Points are observed standardized residuals; dashed lines mark the nominal 95\% prediction interval ($\pm 1.96\sigma$). Reported coverage is the empirical fraction of observations falling within this interval at each isotherm.}
    \label{fig:standardized_residual_isotherms}
\end{figure}

For uncertainty quantification, all partitioned model implementations achieved prediction interval coverage probabilities close to the nominal level. Performance declined when noise was added, as shown by the comparison between $\nu=0$ and $\nu=0.01$ in Table~\ref{tab:band_scores_all}. Figure~\ref{fig:standardized_residual_isotherms} presents standardized prediction residuals and empirical coverage for the Case~2 fit with noise level $\nu=0$, along three representative isotherms spanning the temperature range. Coverage is near-nominal at the lowest and highest temperatures ($T=592$ and $5070$~K, coverage 0.99--1.00) but drops modestly at the intermediate isotherm ($T=1710$~K; 0.87 for $E_{\mathrm{ion}}$, 0.92 for $P_{\mathrm{ion}}$). Complete uncertainty quantification results at the 25th, 50th, and 75th percentiles of temperature and density are provided in the Supplementary Material.

\section{Statistical Modeling of the \texorpdfstring{CO$_2$}{CO2} Equation of State}
\label{sec:casestudy}
We demonstrate the proposed framework on an EOS table for carbon dioxide (CO$_2$). Unlike the simulation study, the true molar mass surface and Debye temperature are unknown. Hence, evaluation focuses on three criteria: (i) fidelity of the reconstructed $E_{\mathrm{ion}}$ and $P_{\mathrm{ion}}$; (ii) physical consistency of the recovered molar mass and Debye-temperature surfaces; and (iii) and the uncertainty quantification provided by the parametric bootstrap.

\subsection{Model Configuration}
\begin{figure}[!htb]
\centering
\includegraphics[width=0.9\textwidth]{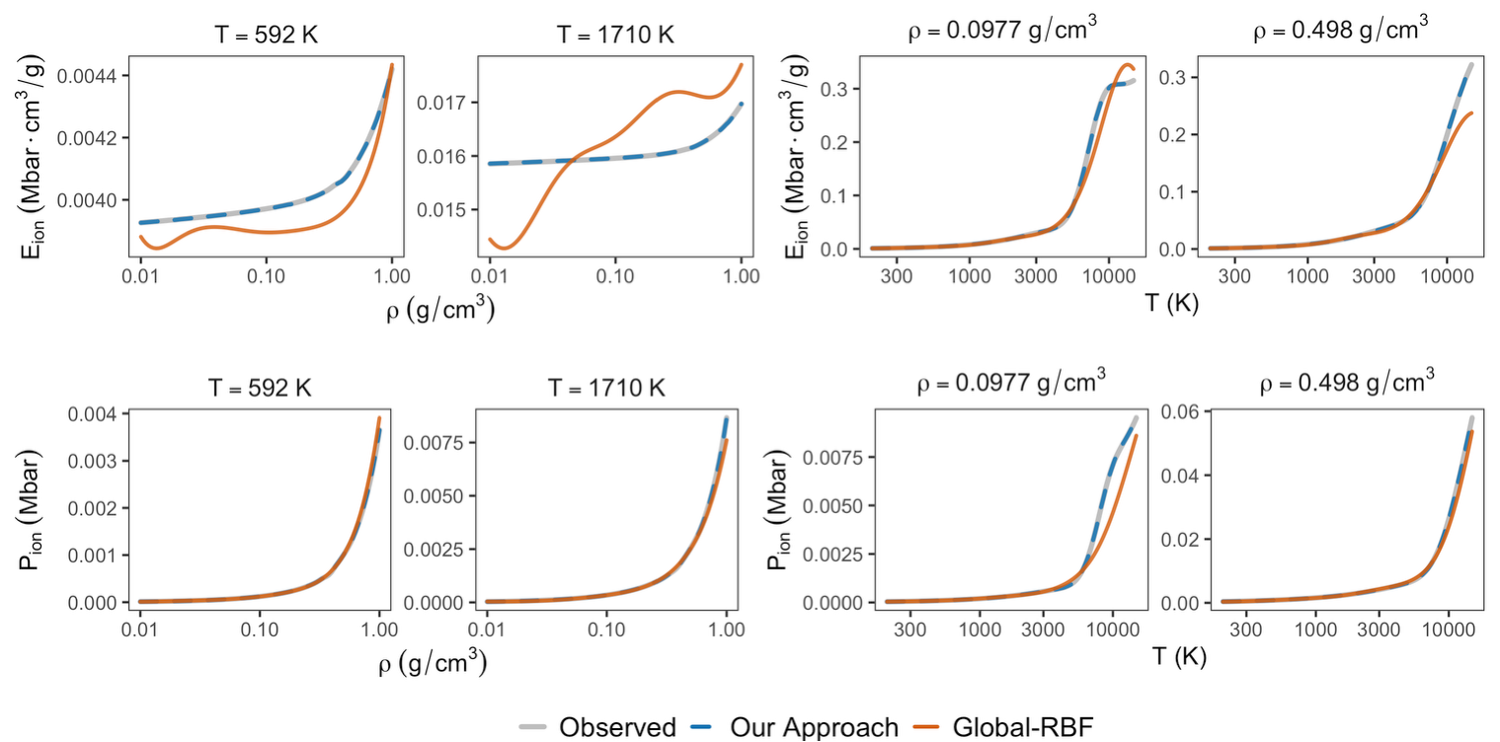}
\caption{Comparison of observed and predicted $E_{\mathrm{ion}}$ (top row) and $P_{\mathrm{ion}}$ (bottom row) for CO$_2$, from our partitioned approach and the Global Cowan--RBF model. The left two columns show isotherms at $T=592$~K and $T=1710$~K, plotted as a function of density $\rho$; the right two columns show isochores at $\rho=0.0977$~g/cm$^3$ and $\rho=0.498$~g/cm$^3$, plotted as a function of temperature $T$.}
\label{fig:compare_CO2_methods}
\end{figure}

The CO$_2$ table contains $N=20{,}000$ points, and we use the same 85\%/15\% training/test split and sampling design as in Section~\ref{sec:design}. Based on the sensitivity analysis described in Section~\ref{subsec:impl} (full results in the Supplementary Material), we set $n_p=20$ for the linear basis and $n_p=15$ for the quadratic and cubic bases, with $\delta=0.01$ throughout; these settings balance low relative error with accurate reconstruction of the isotherms and isochores (Figure~\ref{fig:compare_CO2_methods}). The Global Cowan--RBF comparator uses Gaussian RBFs with fixed bandwidth $0.35$, with the grid of centers selected from several regular grids over the $(x,y)$ domain (Table~\ref{tab:band_scores_test}). All other physical and statistical components are shared, so the comparison isolates the effect of localization.

\subsection{Results}
\begin{figure}[!htb]
    \centering
    \includegraphics[width=0.9\textwidth]{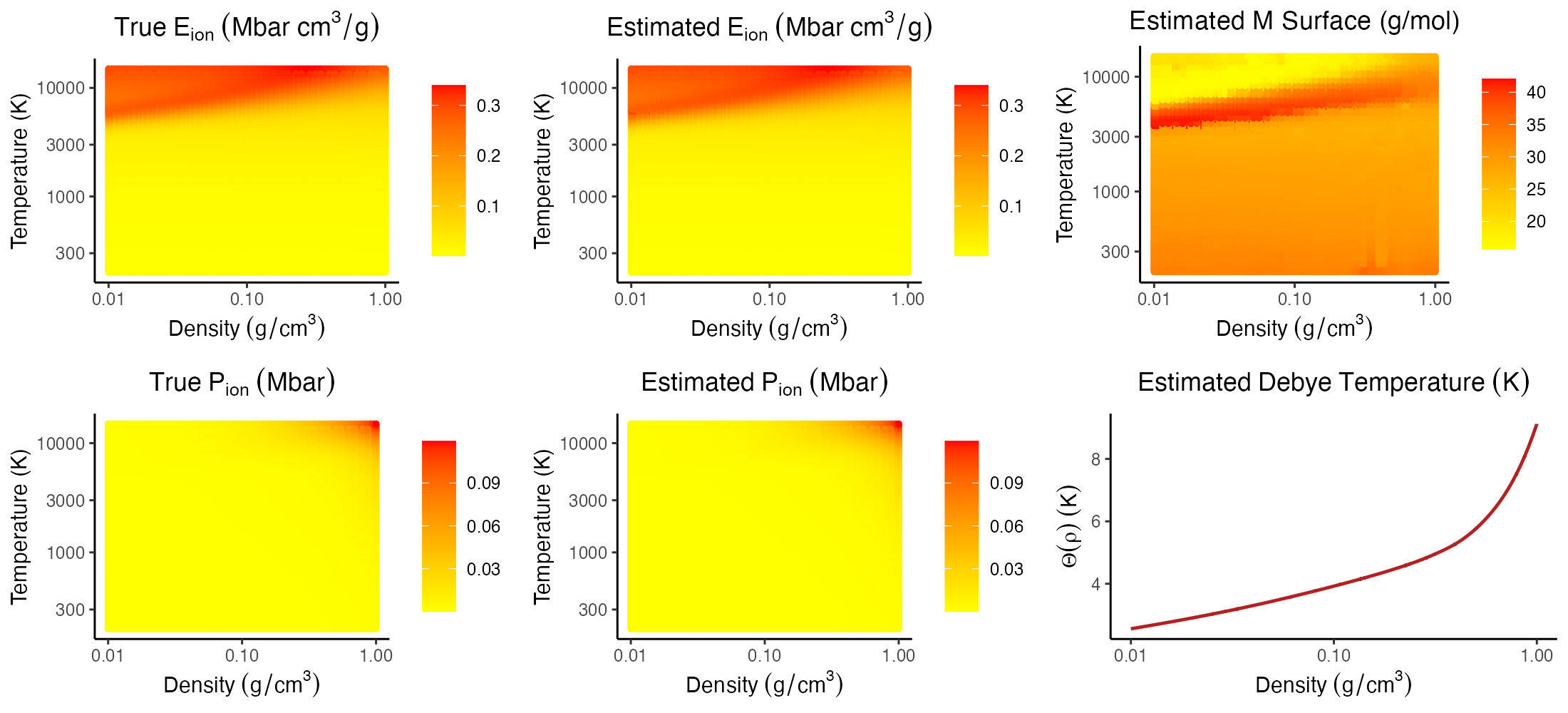}
    \caption{Fitted EOS surfaces for the cubic model ($n_{\text{p}} = 15$, $\delta = 0.01$). Top row: ion
             energy $E_{\text{ion}}$ (Mbar\,cm$^{3}$/g), true and estimated, and
             the estimated molar mass $M$ (g/mol). Bottom row: ion pressure
             $P_{\text{ion}}$ (Mbar), true and estimated, and the estimated Debye
             temperature $\Theta(\rho)$ (K).}
    \label{fig:eos_surfaces_cubic}
  \end{figure}

The cubic basis achieves the lowest held-out relative error loss among the polynomial bases considered (Table~\ref{tab:band_scores_test}), although the gain over the quadratic basis is marginal ($7.90\times10^{-6}$ vs.\ $8.01\times10^{-6}$). There is substantial improvement from the linear to the quadratic bases, which reduces the relative error by roughly an order of magnitude. Across all RBF-center grid sizes, the Global Cowan--RBF model's relative error loss remains at least an order of magnitude larger than the quadratic and cubic partitioned models, even at its best-performing configuration (100 centers, relative error $9.03\times10^{-3}$). Figure~\ref{fig:eos_surfaces_cubic} shows the recovered molar-mass surface $M$ and Debye-temperature curve $\Theta(\rho)$ alongside the fitted $E_{\mathrm{ion}}$ and $P_{\mathrm{ion}}$ surfaces for the cubic model. Both recovered quantities vary smoothly and remain within physically plausible bounds across the density--temperature domain, consistent with the expected dissociation-driven transition in CO$_2$.

\begin{table}[ht]
  \centering
  \caption{Held-out predictive performance and uncertainty-band scores for the
         polynomial models, and test-set relative error loss for the
         competing Global Cowan--RBF method. Empirical coverage is the
         fraction of held-out observations inside the 95\% predictive band
         (nominal 0.95), and scaled interval scores are reported with
         $\alpha=0.05$. Best value in each column is shown in bold.}
  \label{tab:band_scores_test}
  \small

  \textbf{Polynomial models}

  \begin{tabular}{l|c|cc|cc}
    \toprule
     & Relative error & \multicolumn{2}{c|}{95\% predictive coverage}
     & \multicolumn{2}{c}{Interval score (scaled)} \\
    \hline
    Method & loss & $E_{\mathrm{ion}}$ & $P_{\mathrm{ion}}$
           & $E_{\mathrm{ion}}$ & $P_{\mathrm{ion}}$ \\
    \midrule
    Linear
      & 9.60e-05 & 0.931 & \textbf{0.966} & 0.0315 & 0.0229 \\
    Quadratic
      & 8.01e-06 & 0.943 & 0.924 & 0.0075 & 0.0072 \\
    Cubic
      & \textbf{7.90e-06} & \textbf{0.945} & 0.929
      & \textbf{0.0074} & \textbf{0.0071} \\
    \bottomrule
  \end{tabular}

  \vspace{0.8em}

  \textbf{Global Cowan--RBF models}

  \begin{tabular}{l|c}
    \toprule
    Method & Relative error loss \\
    \midrule
    Global Cowan--RBF (16)  & 0.04151443 \\
    Global Cowan--RBF (25)  & 0.02964553 \\
    Global Cowan--RBF (36)  & 0.02358511 \\
    Global Cowan--RBF (64)  & 0.01107595 \\
    Global Cowan--RBF (100) & 0.00903322 \\
    \bottomrule
  \end{tabular}
\end{table}

\begin{figure}[!htb]
    \centering
    \includegraphics[width=0.85\textwidth]{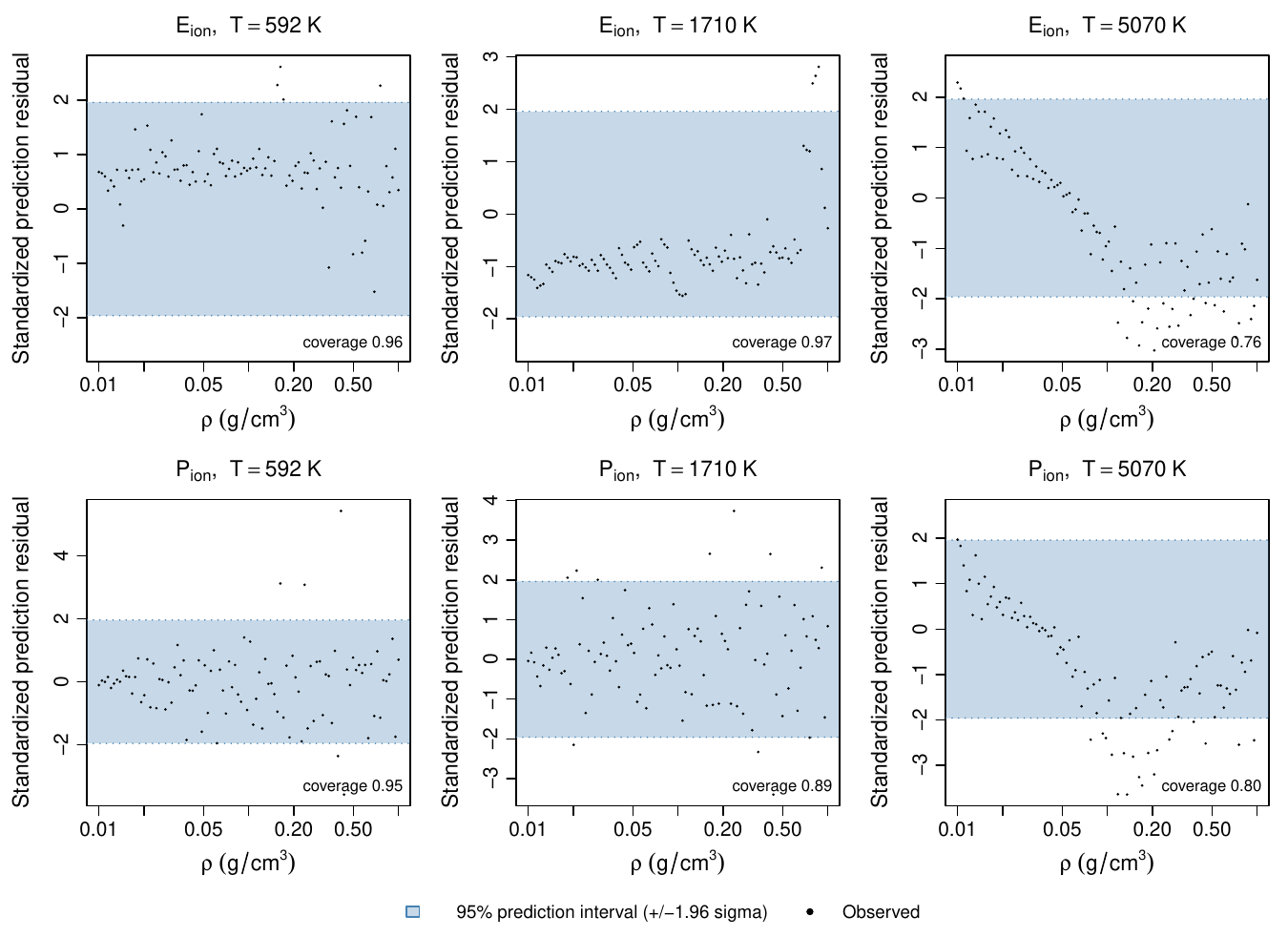}
    \caption{Standardized prediction residuals for $E_{\mathrm{ion}}$ (top row) and $P_{\mathrm{ion}}$ (bottom row) as a function of density $\rho$, for the real CO$_2$ dataset. These are shown along three representative isotherms ($T = 592$, $1710$, and $5070$~K). Points are observed standardized residuals and dashed lines mark the nominal 95\% prediction interval. The reported coverage is the empirical fraction of observations falling within this interval at each isotherm.}
    \label{fig:standardized_residual_isotherms_CO2}
\end{figure}

Since the CO$_2$ table is deterministic, the bootstrap intervals measure the stability of the estimated molar-mass surfaces and Debye temperature under observational error (i.e., residual discrepancy), given the chosen Cowan model. They should not be interpreted as uncertainty in the physical EOS itself or as accounting for uncertainty in the Cowan model form. The cubic basis attains the best interval scores for both $E_{\mathrm{ion}}$ and $P_{\mathrm{ion}}$, with empirical coverage between 0.92 and 0.97 across bases (Table~\ref{tab:band_scores_test}). Along individual isotherms (Figure~\ref{fig:standardized_residual_isotherms_CO2}), coverage is near-nominal at low-to-moderate temperatures but drops at $T=5070$~K (0.76 for $E_{\mathrm{ion}}$, 0.80 for $P_{\mathrm{ion}}$), where the dissociation-driven transition is most pronounced. Overall, the proposed partitioned method recovers the $E_{\mathrm{ion}}$ and $P_{\mathrm{ion}}$ surfaces substantially more accurately than the Global Cowan--RBF method.

\section{Discussion}
\label{sec:discussion}

This paper develops a partitioned semiparametric framework for estimating equation-of-state (EOS) tables. It jointly estimates two constrained latent surfaces: a bounded molar-mass surface $M(\rho,T)$ and a monotone Debye temperature curve $\Theta(\rho)$. Both feed into a shared Cowan free-energy model, so the resulting estimates of ion-thermal energy $E_{\mathrm{ion}}$ and pressure $P_{\mathrm{ion}}$ are thermodynamically consistent by construction. Local polynomial fits for $M$ run in parallel across overlapping partitions of the density--temperature domain, while $\Theta$ is estimated globally with a monotone I-spline; the two are updated jointly through alternating optimization. This setup handles constraints that are common in physical applications but less standard in the semiparametric literature: nonstationary local structure in one surface, monotonicity and boundedness in both, and a nonlinear link to the observed responses.

Across twelve simulated scenarios, the method recovers the true latent surfaces accurately, and the partitioned structure scales estimation to large tables through parallel local fits. In the CO$_2$ case study, where ground truth is unavailable, the method improves held-out predictive accuracy over a physically matched but nonpartitioned Global Cowan--RBF comparator, with the largest gains in the density--temperature regions where CO$_2$ is known to dissociate \citep{wu2019equation}. This supports the value of localized estimation for nonstationary processes and validates the applied motivation for the method: materials like CO$_2$ need EOS parameters that vary with density and temperature rather than a single fixed value \citep{ross1985matter, maerzke2019equations}. The framework is not specific to CO$_2$ and extends naturally to other dissociative materials.

The approach has several limitations. First, because the framework partitions the domain, extrapolation beyond the observed data range within a partition can be unreliable, particularly at higher temperatures and densities with little or no table coverage. This reflects our use of a dense, well-populated EOS table here; as discussed in Section~\ref{sec:inverse_problem}, our ultimate goal is to work directly from raw simulated and experimental data rather than a pre-compiled table, and such data are inherently sparse and irregularly sampled across the density--temperature domain. This sparsity will stress the partitioned structure's reliance on local coverage more than the dense setting considered here. Extending the method into an extrapolation framework, for instance by strategically placing anchor points, expanding boundary partitions, or designing experiments that constrain the sparse regime and reduce the resulting uncertainty, is a natural direction for future work. Second, the recovered surfaces and their uncertainties are conditional on the Cowan free-energy model; if this physical form is misspecified, both $M(\rho,T)$ and $\Theta(\rho)$ absorb the bias. In the deterministic CO$_2$ application, the parametric bootstrap quantifies conditional refitting variability under a working residual-discrepancy model, not model-form or repeated-measurement uncertainty.

Our proposed automated and partitioned approach improves reconstruction accuracy in a real dissociating-fluid application relative to a global alternative, and replaces a process historically driven by manual hand-tuning and material-specific calibration with one that is automated. More broadly, this methodology can be extended to new materials and yields uncertainty estimates that can be propagated into the downstream hydrodynamic simulations, which rely on EOS tables. Future work includes extending the framework from dense tables to sparse simulation and experimental data \citep{wu2019equation}, further characterizing identifiability under alternative structural constraints, and extending the bootstrap to propagate upstream data uncertainty and quantify free-energy model-form uncertainty.

\section*{Declaration of Generative AI Use}
Claude (Anthropic), using the Claude Opus and Sonnet models available during this period, was used between February and September 2026 to assist with software development and manuscript preparation. Its use was limited to helping debug the analysis code, drafting explanatory comments within the code, and editing the authors' original writing for typographical errors and clarity. Claude was not used to design the methodology, conduct the statistical analyses, interpret the results, or generate scientific conclusions. All AI-assisted code changes and comments were reviewed, tested, and verified by the authors.

\if0\blind{%

\section*{Data availability}
The data supporting this study is publicly available in the GitHub repository \url{https://github.com/jolypich/EOS_code.git}.

\section*{Acknowledgments}\label{Sec:Acknowledgements}
This work was performed under the auspices of the U.S. Department of Energy by Lawrence Livermore National Laboratory under Contract DE-AC52-07NA27344.

This document was prepared as an account of work sponsored by an agency of the United States government. Neither the United States government nor Lawrence Livermore National Security, LLC, nor any of their employees makes any warranty, expressed or implied, or assumes any legal liability or responsibility for the accuracy, completeness, or usefulness of any information, apparatus, product, or process disclosed, or represents that its use would not infringe privately owned rights. Reference herein to any specific commercial product, process, or service by trade name, trademark, manufacturer, or otherwise does not necessarily constitute or imply its endorsement, recommendation, or favoring by the United States government or Lawrence Livermore National Security, LLC. The views and opinions of authors expressed herein do not necessarily state or reflect those of the United States government or Lawrence Livermore National Security, LLC, and shall not be used for advertising or product endorsement purposes. \textbf{Release No.}: LLNL-JRNL-2025291.
}\fi

\bibliographystyle{apalike}
\bibliography{main}

\end{document}


\def\spacingset#1{\renewcommand{\baselinestretch}%
{#1}\small\normalsize} \spacingset{1}

\if0\blind
{
\title{\bf Physics-Constrained Modeling and Inference of the Equation of State for Dissociative Materials}

\author{
Jolypich Pek$^{1}$,
Ben Seiyon Lee$^{1,*}$,
Jason Bernstein$^{2}$,
and Philip Myint$^{2}$\\[0.2cm]
$^{1}$Department of Statistics, George Mason University, Fairfax, Virginia, USA\\
$^{2}$Lawrence Livermore National Laboratory, Livermore, California, USA\\[0.1cm]
$^{*}$Corresponding author: \texttt{slee287@gmu.edu}
}

\maketitle
}
 \fi

\if1\blind
{
  \bigskip
  \bigskip
  \bigskip
  \begin{center}
    {\LARGE\bf Physics-Constrained Modeling and Inference of the Equation of State for Dissociative Materials}
\end{center}
  \medskip
} \fi

\spacingset{2} 

\section{Supplementary Material}
The Supplement provides additional details supporting the main text. It includes the full decomposition of the EOS quantities, derivations for the Cowan ion-thermal model, a proof of structural nonidentifiability, details of the polynomial basis functions, parametric bootstrap procedure, and residual-adjusted prediction intervals. It also provides additional results for the CO$_2$ application and simulation study, including the construction of the synthetic molar-mass and Debye-temperature surfaces, comparator models, sensitivity analyses, and results across all simulated datasets.

\subsection{Full Decomposition of Internal Energy and Pressure}
\label{sec:supp_coldelectron}
As introduced in Section~\ref{sec:freeenergy}, the total Helmholtz free energy is decomposed into cold, ionic, and electron contributions,
\begin{equation}
    F(\rho,T) = F_{\mathrm{cold}}(\rho) + F_{\mathrm{ion}}(\rho,T) + F_{\mathrm{electron}}(\rho,T),
\end{equation}
where $F_{\mathrm{cold}}(\rho)$ is the zero-temperature ($T=0$) energy of the static lattice as a function of compression alone, and $F_{\mathrm{electron}}(\rho,T)$ accounts for thermal excitation of the electrons. This decomposition carries over directly to the internal energy and pressure via the thermodynamic identities in Eqs.~\eqref{eq:InternalEnergy}--\eqref{eq:Pressure} of the main text, giving
\begin{align}
    E(\rho,T) &= E_{\mathrm{cold}}(\rho) + E_{\mathrm{ion}}(\rho,T) + E_{\mathrm{electron}}(\rho,T), \\
    P(\rho,T) &= P_{\mathrm{cold}}(\rho) + P_{\mathrm{ion}}(\rho,T) + P_{\mathrm{electron}}(\rho,T),
\end{align}
where each term is obtained by applying Eqs.~\eqref{eq:InternalEnergy} and \eqref{eq:Pressure} to the corresponding free-energy contribution alone. Below we describe the models used for the cold and electron contributions.

\subsection{Cowan-type ion-thermal free energy}
Let $\alpha$ be a material parameter and $R$ the molar gas constant (in the units used by the implementation). Define
\[
T_{\mathrm{melt}}(\rho) = \alpha\,\rho^{-2/3}\,\Theta(\rho)^2,
\qquad
\chi(\rho,T) = \frac{\Theta(\rho)}{T},
\qquad
w(\rho,T) = \frac{T_{\mathrm{melt}}(\rho)}{T}.
\]
With
\[
A(\chi,w) = -\frac{11}{2} + \frac{9}{2}w^{1/3} + \frac{3}{2}\ln\!\left(\frac{\chi^2}{w}\right),
\]
the model is
\begin{equation}\label{eq:Cowan_F}
F_{\mathrm{Cowan}}(\rho,T) = \frac{R\,T}{M(\rho,T)}\,A\!\big(\chi(\rho,T),w(\rho,T)\big).
\end{equation}

\subsection{Energy and pressure from log-derivatives}
Since the table is indexed by $(\log_{10}\rho,\log_{10}T)$, we compute derivatives in log-coordinates and convert them using
\[
\frac{\partial}{\partial T}
=
\frac{1}{(\ln 10)\,T}\frac{\partial}{\partial \log_{10}T},
\qquad
\frac{\partial}{\partial \rho}
=
\frac{1}{(\ln 10)\,\rho}\frac{\partial}{\partial \log_{10}\rho}.
\]
The ion-thermal internal energy and pressure are computed as
\[
E_{\mathrm{ion}}(\rho,T)
=
F_{\mathrm{Cowan}} - \frac{1}{\ln 10}\frac{\partial F_{\mathrm{Cowan}}}{\partial \log_{10}T},
\qquad
P_{\mathrm{ion}}(\rho,T)
=
\frac{\rho}{\ln 10}\frac{\partial F_{\mathrm{Cowan}}}{\partial \log_{10}\rho}.
\]

\subsection{Proof of Proposition~\ref{prop:nonident}}
\begin{proposition}[Structural nonidentifiability]\label{prop:nonident}
In the unrestricted functional model, the Debye temperature curve
$\Theta(\rho)$ and the molar-mass surface $M(\rho,T)$ are not separately
identifiable from observations of $E_{\mathrm{ion}}(\rho,T)$ and
$P_{\mathrm{ion}}(\rho,T)$. For any admissible pair $\{\Theta,M\}$ and any
smooth positive curve $\widetilde\Theta\neq\Theta$, there exists a
molar-mass surface $\widetilde M$ such that the pair
$\{\widetilde\Theta,\widetilde M\}$ reproduces $E_{\mathrm{ion}}$ and
$P_{\mathrm{ion}}$ exactly.
\end{proposition}
\begin{proof}

Let $\mathcal D\subset\mathbb R_+^2$ denote the temperature--density domain and $B(\rho,T;\Theta)=A\big(\chi(\rho,T),w(\rho,T)\big)$ be the bracketed factor of the Cowan free energy \eqref{eq:Cowan_F}, so that
\[
F_{\mathrm{Cowan}}(\rho,T)=RT\,\frac{B(\rho,T;\Theta)}{M(\rho,T)};
\]
the notation $B(\rho,T;\Theta)$ makes the dependence on $\Theta$ explicit. Define the combined ratio $Q(\rho,T)=B(\rho,T;\Theta)/M(\rho,T)$, so that $F_{\mathrm{Cowan}}=RT\,Q$. Applying the thermodynamic identities \eqref{eq:InternalEnergy} and \eqref{eq:Pressure} to $F_{\mathrm{Cowan}}$ gives
\begin{align*}
  E_{\mathrm{ion}}(\rho,T) &= -RT^2\,Q_T(\rho,T),\\
  P_{\mathrm{ion}}(\rho,T) &= \rho^2RT\,Q_\rho(\rho,T),
\end{align*}
where $Q_T$ and $Q_\rho$ denote the partial derivatives of $Q$ with respect to $T$ and $\rho$. Hence $E_{\mathrm{ion}}$ and $P_{\mathrm{ion}}$ depend on $\Theta$ and $M$ only through the partial derivatives of $Q$.

Let $\{\Theta,M\}$ be any admissible pair with $M\neq 0$ and $B(\rho,T;\Theta)\neq 0$ on $\mathcal D$, and let $\widetilde\Theta\neq\Theta$ be any smooth positive curve with $B(\rho,T;\widetilde\Theta)\neq 0$ on $\mathcal D$. Define
\[
  \widetilde M(\rho,T)
  = M(\rho,T)\,
    \frac{B(\rho,T;\widetilde\Theta)}{B(\rho,T;\Theta)}.
\]
Then
\[
  \widetilde Q(\rho,T)
  = \frac{B(\rho,T;\widetilde\Theta)}{\widetilde M(\rho,T)}
  = \frac{B(\rho,T;\widetilde\Theta)}
         {M(\rho,T)\,B(\rho,T;\widetilde\Theta)/B(\rho,T;\Theta)}
  = \frac{B(\rho,T;\Theta)}{M(\rho,T)}
  = Q(\rho,T).
\]
Since $\widetilde Q=Q$ pointwise on $\mathcal D$, we have $\widetilde Q_T=Q_T$ and $\widetilde Q_\rho=Q_\rho$, and therefore
\begin{align*}
  \widetilde E_{\mathrm{ion}}(\rho,T)
    &= -RT^2\,\widetilde Q_T(\rho,T)
     = -RT^2\,Q_T(\rho,T)
     = E_{\mathrm{ion}}(\rho,T),\\
  \widetilde P_{\mathrm{ion}}(\rho,T)
    &= \rho^2RT\,\widetilde Q_\rho(\rho,T)
     = \rho^2RT\,Q_\rho(\rho,T)
     = P_{\mathrm{ion}}(\rho,T).
\end{align*}
Hence the map $\{\Theta,M\}\mapsto\{E_{\mathrm{ion}},P_{\mathrm{ion}}\}$ is not injective, and $\Theta(\rho)$ and $M(\rho,T)$ are not separately identifiable in the unrestricted functional model. This nonidentifiability is structural: it holds for any $\widetilde\Theta\neq\Theta$, regardless of whether $E_{\mathrm{ion}}$ and $P_{\mathrm{ion}}$ are observed with or without noise and irrespective of sample size. Although $\Theta(\rho)$ does not depend on $T$, this does not resolve the nonidentifiability, because the unrestricted molar-mass surface $M(\rho,T)$ can absorb any change in $\Theta$ via the compensator $\widetilde M$.
\end{proof}

\subsection{Polynomial Basis Functions}
Within each partition $\mathcal{K}_h$, we model the partition-specific predictor $\eta_{M,h}(x,y)$ as a polynomial basis expansion in the partition-centered coordinates $(x - x_h, y - y_h)$, including cross terms, where \[ x_h = \frac{1}{|\mathcal{K}_h|}\sum_{i\in \mathcal{K}_h}x_i, \qquad y_h = \frac{1}{|\mathcal{K}_h|}\sum_{i\in \mathcal{K}_h}y_i. \] The partition-specific predictor $\eta_{M,h}(x,y)$ takes one of three polynomial degrees---bilinear (4 parameters), quadratic (6 parameters), or cubic (10 parameters): \begin{equation} \label{eq:eta_M_polynomial_models} \begin{aligned} \mbox{\textit{Bilinear}:}\qquad \eta_{M,h}(x,y) &= a_{0,h} + a_{1,h}(x-x_h) + a_{2,h}(y-y_h) + a_{3,h}(x-x_h)(y-y_h), \\[0.5em] \mbox{\textit{Quadratic}:}\qquad \eta_{M,h}(x,y) &= a_{0,h} + a_{1,h}(x-x_h) + a_{2,h}(y-y_h) \\ &\quad + a_{3,h}(x-x_h)^2 + a_{4,h}(x-x_h)(y-y_h) + a_{5,h}(y-y_h)^2, \\[0.5em] \mbox{\textit{Cubic}:}\qquad \eta_{M,h}(x,y) &= a_{0,h} + a_{1,h}(x-x_h) + a_{2,h}(y-y_h) \\ &\quad + a_{3,h}(x-x_h)(y-y_h) + a_{4,h}(x-x_h)^2 + a_{5,h}(y-y_h)^2 \\ &\quad + a_{6,h}(x-x_h)^3 + a_{7,h}(y-y_h)^3 \\ &\quad + a_{8,h}(x-x_h)^2(y-y_h) + a_{9,h}(x-x_h)(y-y_h)^2. \end{aligned} \end{equation}

\subsection{Parametric Bootstrapping Implementation Details}

In our implementation, bootstrap generation uses a fixed random seed for reproducibility, and each replicate is refit independently using parallel computation across overlapping partitions. We use $B=100$ bootstrap replicates. Increasing $B$ can improve the stability of the estimated bootstrap intervals \citep{efron1994introduction}, though at greater computational cost.
The squared relative residuals are smoothed on the log scale, which guarantees positive variance estimates. For energy and pressure alike, we fit a bivariate local quadratic regression (\texttt{loess} in R, degree $2$, span $0.10$) of $\log(r_{E,i}^2 + \epsilon_E)$ and $\log(r_{P,i}^2 + \epsilon_P)$ on $(x,y)$. The small offsets $\epsilon_E$ and $\epsilon_P$ are the first percentiles of the strictly positive $r_{E,i}^2$ and $r_{P,i}^2$, respectively, and prevent $\log 0$. The relative standard deviations are
\[
\hat{\sigma}_E(x,y) = \sqrt{\exp\!\big\{\widehat{\log(r_{E}^2+\epsilon_E)}(x,y)\big\}},
\qquad
\hat{\sigma}_P(x,y) = \sqrt{\exp\!\big\{\widehat{\log(r_{P}^2+\epsilon_P)}(x,y)\big\}},
\]
each with a floor of $10^{-6}$ for numerical stability. The polynomial degree and span may be adjusted to control the smoothness of the estimated surfaces. The lower empirical quantile levels used to determine $c_E$ and $c_P$ are both set to $0.01$. These levels may be adjusted on a case-by-case basis.

\subsection{Residual-Adjusted Prediction Bands}\label{sec:pi}
The prediction band combines two components under the working residual model: variation in the refitted surface, obtained by parametric bootstrap, and local residual discrepancy, obtained from a smoothed model of the baseline residuals. In the simulation study, the latter corresponds to the injected data-generating noise. For the deterministic CO$_2$ table, it instead represents discrepancies between the fitted model and tabulated values and should not be interpreted as repeated-measurement noise. The residual model is fitted once from the baseline, used to generate the bootstrap replicates, and then re-used as the second variance component, so the same fitted local residual standard deviations are used in both roles. The two variance components are added in quadrature and the band is centered on the bootstrap median. All coverage statements are therefore conditional on this working residual model and on the assumed Cowan free-energy representation.

  For $b=1,\dots,B$ with $B=100$, synthetic data are drawn at the training
  coordinates and the full estimator is rerun on each,
  \begin{equation}
    E_{\mathrm{ion},i}^{(b)}=\hat{E}_{\mathrm{ion},i}+\sigma_{E,i} Z_{E,i}^{(b)},\quad
    Z_{E,i}^{(b)}\stackrel{\text{iid}}{\sim}\mathcal N(0,1),
    \label{eq:boot}
  \end{equation}
  where $\sigma_{E,i} = s_{E,i}\,\hat{\sigma}_E(x_i,y_i)$ is the local
  absolute noise standard deviation, with $s_{E,i}$ and $\hat{\sigma}_E$ as defined
  in Section~\ref{sec:bootstrap}, giving replicate surfaces evaluated at every $(x_i,y_i)$. With
  $\tilde{E}_{\mathrm{ion},i}=\operatorname{med}_b \hat{E}_{\mathrm{ion},i}^{(b)}$ and
  $\hat{s}_{E,i}=\operatorname{sd}_b \hat{E}_{\mathrm{ion},i}^{(b)}$, the $95\%$ prediction interval is
  \begin{equation}
    \tilde{E}_{\mathrm{ion},i} \;\pm\; z_{0.975}\sqrt{\hat{s}_{E,i}^{2}+\sigma_{E,i}^{2}},
    \qquad z_{0.975}=\Phi^{-1}(0.975),
    \label{eq:pi}
  \end{equation}
  where $\hat{s}_{E,i}^{2}$ is refitting variability and $\sigma_{E,i}^{2}$ is the local residual-discrepancy variance assigned to a held-out table value. The residual term is Gaussian by construction
  \eqref{eq:boot}; the estimation term is treated as Gaussian and the two as
  independent. At held-out coordinates $\sigma_{E,i}$ is evaluated using the
  smoother $\hat{\sigma}_E$ from Section~\ref{sec:bootstrap}, which is defined
  off the training grid, so no held-out observation enters its own interval.
  The identical construction applies to $P_{\mathrm{ion}}$, with $\sigma_{E,i}$,
  $\hat{\sigma}_E$, $s_{E,i}$, and $Z_{E,i}^{(b)}$ replaced by their
  $P$-counterparts throughout.

\subsubsection{Standardized Prediction Deviations}
The isotherm and isochore plots can be difficult to interpret because the predicted and observed values of $E_{\mathrm{Ion}}$ and $P_{\mathrm{Ion}}$ are very similar and the corresponding prediction intervals are narrow relative to the scale of the response. To improve visual inspection of the deviations and prediction intervals, we instead plot the standardized prediction deviations
\[
u = \frac{z-c}{s},
\]
where $z$ is the observed value, $c$ is the center of the 95\% prediction interval, and $s$ is the total predictive standard deviation. Since the prediction interval is constructed as $c \pm z_{0.975}s$
where $z_{0.975}=\Phi^{-1}(0.975)\approx 1.96$, the predictive standard deviation can be recovered from the interval limits as
\[
s
=
\frac{U-L}{2z_{0.975}}
=
\frac{U-L}{2\times 1.96},
\]
where $L$ and $U$ denote the lower and upper limits of the 95\% prediction interval, respectively. On this scale, the prediction interval corresponds to $-1.96 \leq u \leq 1.96$
so values outside this range indicate prediction interval coverage failures.
Suppose the prediction interval is
\[
[L,U]
=
E_{\mathrm{pred}} \pm z_{0.975}s,
\]
where \(z_{0.975}\approx 1.96\). Its total width is $
U-L
=
2(1.96)s$. 
Solving for \(s\) yields
\[
s
=
\frac{U-L}{2z_{0.975}}
.
\]

\subsubsection{Interval Metrics }
In addition to point estimation, we assess the quality of the prediction intervals \eqref{eq:pi} using the prediction interval coverage probability (PICP) \citep{khosravi2010lower, pearce2018high} and the interval score \citep{gneiting2007strictly}. The PICP is the empirical coverage probability that the held out validation data are covered by the derived prediction intervals. These are defined as 
\begin{equation}
\operatorname{PICP}_{E}
=
\frac{1}{n_{\mathrm{pred}}}
\sum_{i=1}^{n_{\mathrm{pred}}}
\mathbb{I}
\left(
L_{E,i}
\leq
E_{\mathrm{ion},i}
\leq
U_{E,i}
\right),
\end{equation}

\begin{equation}
\operatorname{PICP}_{P}
=
\frac{1}{n_{\mathrm{pred}}}
\sum_{i=1}^{n_{\mathrm{pred}}}
\mathbb{I}
\left(
L_{P,i}
\leq
P_{\mathrm{ion},i}
\leq
U_{P,i}
\right),
\end{equation}

We assess interval estimation performance with the interval score \citep{gneiting2007strictly}. For a nominal $(1-\alpha)$ interval $[\ell_i,u_i]$ and observation $z_i$,
\begin{equation}
S_\alpha(\ell_i,u_i;z_i) = (u_i-\ell_i)
 + \tfrac{2}{\alpha}(\ell_i-z_i)\mathbf{1}\{z_i<\ell_i\}
 + \tfrac{2}{\alpha}(z_i-u_i)\mathbf{1}\{z_i>u_i\},
\label{eq:interval_score}
\end{equation}
with $\alpha=0.05$. We report the mean interval score across locations, with smaller values indicating narrower intervals that cover the true values.

\subsection{\texorpdfstring{CO$_2$}{CO2} Equation of State Application}
\begin{table}[h]
  \centering
  \caption{Held-out relative-error loss and wall-clock time for the CO$_2$ EOS application across target partition sizes $n_{\mathrm{pp}}$, overlap widths $\delta$, and local polynomial models. Lower loss indicates better predictive performance. Wall-clock times are reported in seconds for the outer iteration loop; each run used 20 processor cores on a single compute node.}
  \label{tab:grid_results}
  \label{tab:grid_walltime}
  \small
  \begin{tabular}{rr|ccc|rrr}
    \toprule
    & & \multicolumn{3}{c|}{Relative-error loss} & \multicolumn{3}{c}{Wall-clock time (s)} \\
    \cmidrule(lr){3-5}\cmidrule(lr){6-8}
    $n_{\mathrm{pp}}$ & $\delta$ & Linear & Quadratic & Cubic & Linear & Quadratic & Cubic \\
    \midrule
     10 & 0.01 & 5.26e-05 & \textbf{5.87e-06} & \textbf{5.85e-06} & 583 & 384 &  255 \\
     15 & 0.01 & 8.19e-05 & 8.01e-06 & \textbf{7.94e-06}          & 311 & 441 &  375 \\
     20 & 0.01 & 9.60e-05 & 1.76e-05 & 1.61e-05                   & 163 & 111 &  161 \\
     25 & 0.01 & 1.34e-04 & 7.27e-05 & 3.70e-05                   & 202 & 103 &  192 \\
     50 & 0.01 & 2.95e-04 & 6.05e-05 & 3.13e-05                   & 155 & 132 &   89 \\
    100 & 0.01 & 5.33e-04 & 1.84e-04 & 1.76e-05                   & 677 &  73 &  138 \\
    \midrule
     10 & 0.05 & 1.51e-04 & 6.71e-05 & 3.34e-05                   & 113 & 112 &  281 \\
     15 & 0.05 & 6.09e-04 & 1.31e-04 & 1.34e-04                   &  80 &  80 & 1032 \\
     20 & 0.05 & 6.76e-04 & 7.83e-04 & 8.71e-04                   & 210 & 129 & 1859 \\
     25 & 0.05 & 4.29e-04 & 5.42e-04 & 4.15e-04                   &  76 & 108 & 2177 \\
     50 & 0.05 & 1.41e-03 & 3.98e-03 & 2.81e-03                   &  71 &  62 & 1605 \\
    100 & 0.05 & 2.48e-03 & 5.58e-03 & 7.44e-03                   &  71 & 101 &  176 \\
    \midrule
     10 & 0.10 & 2.77e-04 & 2.49e-04 & 4.04e-04                   & 255 & 364 & 283 \\
     15 & 0.10 & 1.01e-03 & 1.30e-04 & 1.30e-04                   & 157 & 114 & 137 \\
     20 & 0.10 & 8.66e-04 & 9.93e-04 & 2.26e-03                   & 222 & 107 & 253 \\
     25 & 0.10 & 1.50e-03 & 8.50e-04 & 1.16e-03                   & 191 & 324 & 122 \\
     50 & 0.10 & 3.48e-03 & 5.61e-03 & 7.62e-03                   &  85 & 195 &  85 \\
    100 & 0.10 & 3.48e-03 & 4.92e-03 & 5.22e-03                   & 135 &  76 & 136 \\
    \bottomrule
  \end{tabular}
\end{table}

\subsection{Simulation Study Design Details}

In this subsection, we provide additional details pertaining to the simulation study including construction of the molar mass surfaces, Debye temperature, the Global Cowan--RBF model, and the direct semi-parametric modeling method. 

\subsubsection{Synthetic Molar Mass Surfaces and Noise Model}
\label{sec:supp_synthetic_cases}
This section provides additional detail on the simulation study design described in Section~\ref{sec:design} of the main text: the full list of twelve baseline molar mass forms and the noise model used to generate the synthetic EOS tables.

Table~\ref{tab:synthetic_cases_supp} lists the twelve baseline surfaces $g(x,y)$ underlying the molar mass cases shown in Figure~\ref{fig:molarmass} (main text). Each form is governed by a scale parameter $s$ and shift parameter $b$; since $g(x,y)$ is subsequently mapped through the bounded rescaling in Equation~\eqref{eq:sim_M_rescale}, the specific values of $s$ and $b$ do not affect the resulting molar mass surface, only the case type does, and we fix $s=1$, $b=0$ throughout.

\begin{table}[h]
\centering
\begin{tabular}{|c|l|p{0.45\textwidth}|}
\hline
\textbf{Case} & \textbf{Description} & \textbf{Equation} \\
\hline
1  & $M$ increases with $x$ & $g(x,y)=sx+b$ \\
2  & $M$ decreases with $x$ & $g(x,y)=-sx+b$ \\
3  & $M$ increases with $y$ & $g(x,y)=sy+b$ \\
4  & $M$ decreases with $y$ & $g(x,y)=-sy+b$ \\
5  & $M$ increases with both $x$ and $y$ & $g(x,y)=s(x+y)+b$ \\
6  & $M$ decreases with both $x$ and $y$ & $g(x,y)=-s(x+y)+b$ \\
7  & $M$ increases with $y$, decreases with $x$ & $g(x,y)=s(y-x)+b$ \\
8  & $M$ increases with $x$, decreases with $y$ & $g(x,y)=s(x-y)+b$ \\
9  & Parabolic in $x$ & $g(x,y)=s(x-\mu_x)^2+b$, \ $\mu_x=\mathrm{mean}(x)$ \\
10 & Parabolic in $y$ & $g(x,y)=s(y-\mu_y)^2+b$, \ $\mu_y=\mathrm{mean}(y)$ \\
11 & Sinusoidal in $x$ & $g(x,y)=s\sin\!\left(2\pi\frac{x-x_{\min}}{x_{\max}-x_{\min}}\right)+b$ \\
12 & Mixed sine/cosine basis & $g(x,y)=s\left[\sin(2\pi x)+\cos(2\pi y)\right]+b$ \\
\hline
\end{tabular}
\caption{Baseline surfaces $g(x,y)$ for the twelve synthetic molar mass cases used in the simulation study.}
\label{tab:synthetic_cases_supp}
\end{table}

The simulated $M(x,y)$ Figure~\ref{fig:molarmass} and $\Theta(x)$ Figure~\ref{fig:debye_sim} feed into the Cowan free energy \eqref{eq:Cowan} ($\alpha=2$, $R=8.3145\times10^{-5}$ MBar$\cdot$cm$^3$/mol/K), returning $E_{\mathrm{ion}}$, $P_{\mathrm{ion}}$, and $S_{\mathrm{ion}}$ in closed form. Only $E_{\mathrm{ion}}$ and $P_{\mathrm{ion}}$ are used for fitting; each is perturbed with magnitude-proportional Gaussian noise,
\begin{equation}
\begin{aligned}
\tilde E_{\mathrm{ion},i} &= E_{\mathrm{ion},i} + \varepsilon^{E}_i,
&\varepsilon^{E}_i &\overset{\text{ind}}{\sim} \mathcal{N}\!\Big(0,\ \nu^2\,\max\!\big(|E_{\mathrm{ion},i}|,\,c_{E}\big)^2\Big),\\
\tilde P_{\mathrm{ion},i} &= P_{\mathrm{ion},i} + \varepsilon^{P}_i,
&\varepsilon^{P}_i &\overset{\text{ind}}{\sim} \mathcal{N}\!\Big(0,\ \nu^2\,\max\!\big(|P_{\mathrm{ion},i}|,\,c_{P}\big)^2\Big),
\end{aligned}
\label{eq:sim_noise}
\end{equation}
where $\nu\in\{0,0.01\}$ (noiseless and $1\%$ noise) and the floors $c_E,c_P$ are the first percentiles of $\{|E_{\mathrm{ion},i}|\}$ and $\{|P_{\mathrm{ion},i}|\}$, preventing vanishing variance near zero signal. Crossing noise levels with the twelve surfaces yields $12\times2=24$ EOS tables.



\subsubsection{Debye Temperature Estimation}
The Debye temperature $\Theta(x)$ is constructed from a cubic I-spline expansion in $x$,
\begin{equation}
\eta_\Theta(x) = \beta_0 + \sum_{j=1}^{J}\beta_j\, I_j(x),
\qquad
\Theta(x) = \Theta_{\min} + (\Theta_{\max}-\Theta_{\min})\,
\frac{e^{\eta_\Theta(x)}-\min_{x} e^{\eta_\Theta}}{\max_{x} e^{\eta_\Theta}-\min_{x} e^{\eta_\Theta}},
\label{eq:sim_debye}
\end{equation}
where $\{I_j\}_{j=1}^{J}$ are cubic I-spline basis functions with knots at density quantiles and the coefficients satisfy $\beta_j\ge 0$ for $j\ge 1$, drawn once at random. The nonnegativity constraint makes $\eta_\Theta$ monotone nondecreasing, the exponential enforces positivity, and the affine map fixes the range to $(\Theta_{\min},\Theta_{\max})=(300,5000)$~K.

\begin{figure}[h]
    \centering
    \includegraphics[width=0.6\textwidth]{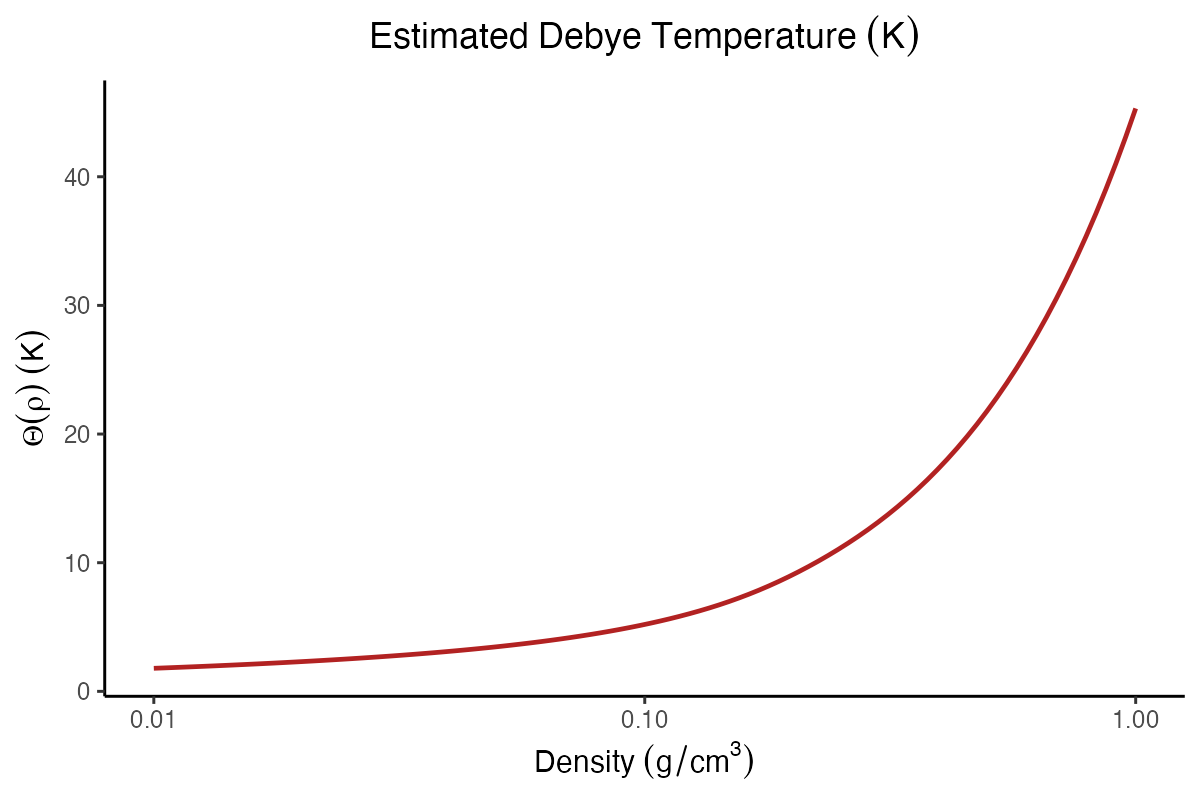}
    \caption{Fitted Debye temperature $\Theta(\rho)$ as a function of density, obtained from a natural cubic spline fit to the tabulated Debye temperature data. This curve is shared across all simulated molar mass cases.}
    \label{fig:debye_temp}
\end{figure}

\subsubsection{Global Cowan--RBF model}
The Global Cowan--RBF model is a physically matched, nonpartitioned comparator: it retains the Cowan free energy, bounded molar-mass link, monotone Debye-temperature curve, joint relative-error loss, and alternating estimation scheme used by the proposed method. Its only structural change is that the molar-mass surface is approximated by Gaussian radial basis functions (RBFs) defined over the entire $(x,y)$
domain, with no spatial partitioning or blending. Centers $\{(c_{x k}, c_{y k})\}_{k=1}^{K}$
are placed on a regular $n_\rho \times n_y$ grid spanning the observed range of
$x=\log_{10}\rho$ and $y = \log_{10}T$, and each basis function is
\begin{equation}
\phi_k(x,y) = \exp\!\left(-\frac{(x-c_{x k})^2}{2\ell_x^2}
-\frac{(y-c_{y k})^2}{2\ell_y^2}\right), \qquad k = 1,\dots,K,
\label{eq:global_rbf_basis}
\end{equation}
with fixed length-scales $\ell_x$ and $\ell_y$ controlling the smoothness of each
bump. The latent surface is a linear combination of these global basis functions,
\begin{equation}
\eta_M(x,y) = \delta_0 + \sum_{k=1}^{K} \delta_k\, \phi_k(x,y),
\label{eq:global_rbf_eta}
\end{equation}
and is mapped to the physical molar-mass range through the same bounded sigmoid link
used elsewhere in the model,
\begin{equation}
M(x,y) = M_{\min} + (M_{\max}-M_{\min})\,\sigma\!\big(\eta_M(x,y)\big),
\qquad \sigma(u) = \frac{1}{1+e^{-u}}.
\label{eq:global_rbf_M}
\end{equation}
Unlike the partitioned approach, a single coefficient vector $\boldsymbol\delta =
(\delta_0,\dots,\delta_K)$ governs the surface everywhere, so local features must be
captured by superposing global bumps rather than by dedicated local basis expansions
with continuity constraints at region boundaries. The coefficients are fit by
penalized least squares,
\begin{equation}
\hat{\boldsymbol\delta} = \arg\min_{\boldsymbol\delta}\;
\mathcal{L}\big(\tilde E_{\mathrm{ion}}, \tilde P_{\mathrm{ion}} \,\big|\, M(x,y;\boldsymbol\delta), \Theta(x)\big)
+ \lambda_M \sum_{k=1}^{K} \delta_k^2,
\label{eq:global_rbf_obj}
\end{equation}
where $\mathcal{L}$ is the same relative-error loss used for the partitioned model and
the ridge penalty $\lambda_M$ shrinks the basis coefficients (excluding the intercept)
to discourage overfitting given the global, unconstrained placement of centers. As in
the proposed method, $\boldsymbol\delta$ and the Debye-temperature spline coefficients
$\boldsymbol\beta$ are updated via alternating optimization until convergence, so the
two approaches differ only in how the latent molar-mass surface is represented, isolating
the effect of partitioning from the rest of the estimation procedure.

\subsubsection{Direct (Semiparametric) Modeling of EOS tables}
As a third comparison method, we also fit $E_{\mathrm{ion}}$ and
$P_{\mathrm{ion}}$ directly as smooth functions of $(x,y)$, bypassing the Cowan
free-energy structure altogether. This approach does not posit a latent molar-mass
surface $M(x,y)$ or Debye-temperature curve $\Theta(x)$ at all; instead each
thermodynamic quantity is modeled as its own black-box response surface, fit
independently of the other. Concretely, we fit two separate generalized additive
models (GAMs) with a tensor-product smooth in $(x,y)$ and a Gamma family with
log link, reflecting the strictly positive, heteroscedastic nature of $E_{\mathrm{ion}}$
and $P_{\mathrm{ion}}$,
\begin{equation}
\log \mathbb{E}\big[\tilde E_{\mathrm{ion}} \mid x,y\big] = f_E(x,y),
\qquad
\log \mathbb{E}\big[\tilde P_{\mathrm{ion}} \mid x,y\big] = f_P(x,y),
\label{eq:direct_gam}
\end{equation}
where $f_E$ and $f_P$ are each represented by a tensor-product spline basis (thin-plate
marginal bases in $x$ and $y$) with $20 \times 20$ basis dimensions, and the
smoothing parameters are chosen by restricted maximum likelihood (REML). Since
$f_E$ and $f_P$ are estimated independently and unconstrained by any shared
free-energy potential, this model can reproduce the observed $E_{\mathrm{ion}}$ and
$P_{\mathrm{ion}}$ tables arbitrarily well given enough flexibility, but by
construction it cannot recover a latent molar-mass surface or Debye-temperature curve,
and nothing enforces thermodynamic consistency between the fitted $E_{\mathrm{ion}}$
and $P_{\mathrm{ion}}$ surfaces. This baseline therefore represents the intuitive,
``fit whatever surface matches the data'' alternative, and serves to demonstrate that
high reconstruction fidelity of the observed table alone is insufficient: a model can
match $E_{\mathrm{ion}}$ and $P_{\mathrm{ion}}$ closely while providing no
insight into, or even contradicting, the underlying molar-mass and Debye-temperature
structure that the proposed method targets directly.

\subsection{Simulation Study Results}
In this section, we present additional simulation results omitted from the main manuscript, including sensitivity analyses and relevant plots for all 24 datasets.
 
\subsubsection*{Sensitivity analysis across simulated datasets.}
This subsection presents the full sensitivity analysis for all 24 simulated datasets using relative-error loss on the held-out test set. We compare results across partition sizes $n_p \in \{10,15,20,25,50,100\}$, overlap fractions $\delta \in \{0.01,0.05,0.10\}$, polynomial models (linear, quadratic, and cubic), and relative noise standard deviations $\nu \in \{0,0.01\}$.

\input{grid_results_all.tex}

\subsubsection*{Global radial basis functions modeling results across simulated datasets.}
This subsection presents the results for the radial basis functions approach. We provide the relative-error loss on the held-out test set of all 24 simulated datasets. Each RBF implementation utilizes $k = 16, 25$ basis functions with equally-spaced knot locations on the temperature-density domain.

\input{global_rbf_simulated_error_table.tex}
\bibliographystyle{apalike}
\bibliography{main}

%% file: bootstrap_band_scores_all.tex
%

\begin{table}[ht]
  \centering
  \caption{Held-out predictive and uncertainty metrics for simulated datasets generated under two relative noise standard deviations, $\nu\in\{0,0.01\}$, and fit using the selected polynomial basis and partition configuration. The metrics include the relative error loss and the scaled interval score ($\alpha=0.05$) and prediction interval coverage probability (PICP).}
  \label{tab:band_scores_all}
  \small
  \begin{tabular}{rrlr|cc|cc|cc}
    \toprule
    & Noise & \multicolumn{2}{c|}{Selected settings}
    & \multicolumn{2}{c|}{Relative Error Loss}
    & \multicolumn{2}{c|}{95\% PICP}
    & \multicolumn{2}{c}{Interval score (scaled)} \\
    \cmidrule(lr){2-2}
    \cmidrule(lr){3-4}
    \cmidrule(lr){5-6}
    \cmidrule(lr){7-8}
    \cmidrule(lr){9-10}
    Case & $\nu$ & Basis & $n_{pp}$
    & Partitioned & Global RBF
    & Energy & Pressure
    & Energy & Pressure \\
    \midrule
     1 & 0 & Quadratic & 15 & 1.80e-06 & 0.027262 & 0.963 & 0.891 & 0.0020 & 0.0042 \\
     1 & 0.01 & Cubic & 50 & 2.07e-04 & 0.027400 & 0.942 & 0.945 & 0.0486 & 0.0470 \\
    \hline
     2 & 0 & Cubic & 15 & 1.00e-05 & 0.116422 & 0.943 & 0.950 & 0.0028 & 0.0082 \\
     2 & 0.01 & Cubic & 100 & 2.14e-04 & 0.116743 & 0.950 & 0.944 & 0.0477 & 0.0506 \\
    \hline
     3 & 0 & Cubic & 20 & 6.19e-06 & 0.261430 & 0.939 & 0.953 & 0.0083 & 0.0036 \\
     3 & 0.01 & Cubic & 100 & 2.02e-04 & 0.260995 & 0.941 & 0.941 & 0.0474 & 0.0485 \\
    \hline
     4 & 0 & Cubic & 100 & 1.45e-05 & 0.035133 & 0.954 & 0.968 & 0.0111 & 0.0066 \\
     4 & 0.01 & Cubic & 100 & 2.12e-04 & 0.035309 & 0.947 & 0.942 & 0.0483 & 0.0476 \\
    \hline
     5 & 0 & Cubic & 15 & 5.83e-06 & 0.787951 & 0.902 & 0.905 & 0.0089 & 0.0097 \\
     5 & 0.01 & Cubic & 100 & 1.96e-04 & 0.790087 & 0.958 & 0.954 & 0.0458 & 0.0459 \\
    \hline
     6 & 0 & Cubic & 100 & 2.40e-03 & 0.303817 & 0.972 & 0.960 & 0.0614 & 0.0934 \\
     6 & 0.01 & Quadratic & 100 & 1.06e-02 & 0.301597 & 0.959 & 0.954 & 0.1067 & 0.3433 \\
    \hline
     7 & 0 & Cubic & 15 & 1.25e-04 & 4.126757 & 0.941 & 0.906 & 0.0178 & 0.0418 \\
     7 & 0.01 & Cubic & 15 & 5.38e-04 & 4.119084 & 0.932 & 0.925 & 0.0592 & 0.0875 \\
    \hline
     8 & 0 & Linear & 100 & 2.21e-04 & 0.013246 & 0.981 & 0.981 & 0.0342 & 0.0378 \\
     8 & 0.01 & Linear & 100 & 4.25e-04 & 0.013447 & 0.951 & 0.960 & 0.0634 & 0.0612 \\
    \hline
     9 & 0 & Cubic & 50 & 9.68e-05 & 0.135378 & 0.936 & 0.961 & 0.0134 & 0.0293 \\
     9 & 0.01 & Cubic & 100 & 2.73e-04 & 0.135509 & 0.947 & 0.951 & 0.0503 & 0.0551 \\
    \hline
    10 & 0 & Cubic & 15 & 1.40e-05 & 0.206467 & 0.932 & 0.959 & 0.0164 & 0.0038 \\
    10 & 0.01 & Cubic & 100 & 2.29e-04 & 0.206381 & 0.950 & 0.951 & 0.0510 & 0.0475 \\
    \hline
    11 & 0 & Cubic & 50 & 1.52e-04 & 0.246637 & 0.927 & 0.867 & 0.0239 & 0.0582 \\
    11 & 0.01 & Cubic & 15 & 3.44e-04 & 0.246705 & 0.929 & 0.926 & 0.0562 & 0.0680 \\
    \hline
    12 & 0 & Quadratic & 15 & 1.96e-04 & 1.101044 & 0.921 & 0.954 & 0.0484 & 0.0487 \\
    12 & 0.01 & Cubic & 15 & 3.78e-04 & 1.101250 & 0.925 & 0.928 & 0.0656 & 0.0646 \\
    \bottomrule
  \end{tabular}
\end{table}

%% file: grid_results_all.tex


\begin{table}[htbp]
  \centering
  \caption{Held-out test loss by partition size, overlap fraction, and spline
           model (lower is better), at both noise levels. M-surface \texttt{inc\_rho}:
           dataset 1 ($\nu = 0$) and dataset 13 ($\nu = 0.01$).}
  \label{tab:grid_results_inc_rho}
  \small
  \begin{tabular}{rr|ccc|ccc}
    \toprule
    & & \multicolumn{3}{c|}{$\nu = 0$} & \multicolumn{3}{c}{$\nu = 0.01$} \\
    \cmidrule(lr){3-5} \cmidrule(lr){6-8}
    $n_{\text{pp}}$ & $f_{\text{ov}}$ & Linear & Cubic & Quadratic & Linear & Cubic & Quadratic \\
    \midrule
     10 & 0.01 & 2.78e-05 & \textbf{1.28e-06} & \textbf{1.61e-06} & 2.76e-04 & 2.63e-04 & 2.63e-04 \\
     15 & 0.01 & 6.43e-05 & 1.87e-06 & \textbf{1.80e-06} & 2.90e-04 & 2.34e-04 & 2.34e-04 \\
     20 & 0.01 & 5.76e-05 & 2.21e-05 & 2.27e-05 & 2.62e-04 & 2.39e-04 & 2.40e-04 \\
     25 & 0.01 & 6.71e-05 & 2.26e-05 & 2.31e-05 & 2.75e-04 & 2.32e-04 & 2.33e-04 \\
     50 & 0.01 & 6.72e-05 & 1.22e-05 & 3.04e-05 & 2.68e-04 & \textbf{2.07e-04} & 2.31e-04 \\
    100 & 0.01 & 6.67e-05 & 1.59e-05 & 2.05e-05 & 2.93e-04 & 2.15e-04 & 2.16e-04 \\
    \midrule
     10 & 0.05 & 3.21e-05 & 1.18e-05 & 1.85e-06 & 2.86e-04 & 3.30e-04 & 2.48e-04 \\
     15 & 0.05 & 2.33e-04 & 1.57e-05 & 2.16e-05 & 2.51e-04 & 3.07e-04 & 2.40e-04 \\
     20 & 0.05 & 4.69e-05 & 3.77e-05 & 2.61e-05 & 2.50e-04 & 2.32e-04 & 2.36e-04 \\
     25 & 0.05 & 5.54e-05 & 2.14e-05 & 2.22e-05 & 2.60e-04 & 2.30e-04 & 2.30e-04 \\
     50 & 0.05 & 5.57e-05 & 1.68e-04 & 8.63e-06 & 2.45e-04 & 3.41e-04 & 2.12e-04 \\
    100 & 0.05 & 8.92e-05 & 1.50e-05 & 1.42e-05 & 2.88e-04 & 2.41e-04 & \textbf{2.08e-04} \\
    \midrule
     10 & 0.10 & 3.45e-05 & 2.28e-05 & 2.58e-05 & 2.96e-04 & 2.76e-04 & 2.44e-04 \\
     15 & 0.10 & 1.39e-05 & 2.21e-05 & 6.58e-06 & 2.22e-04 & 2.73e-04 & 2.28e-04 \\
     20 & 0.10 & 2.37e-04 & 7.50e-05 & 1.97e-05 & 2.27e-04 & 3.10e-04 & 2.44e-04 \\
     25 & 0.10 & 2.19e-05 & 9.34e-06 & 4.34e-06 & 2.17e-04 & 2.48e-04 & 2.11e-04 \\
     50 & 0.10 & 3.91e-05 & 1.61e-04 & 1.66e-05 & 2.45e-04 & 4.50e-04 & \textbf{2.07e-04} \\
    100 & 0.10 & 7.31e-05 & 1.57e-05 & 1.57e-05 & 2.64e-04 & 2.29e-04 & 2.09e-04 \\
    \bottomrule
  \end{tabular}
\end{table}


\begin{table}[htbp]
  \centering
  \caption{Held-out test loss by partition size, overlap fraction, and spline
           model (lower is better), at both noise levels. M-surface \texttt{dec\_rho}:
           dataset 2 ($\nu = 0$) and dataset 14 ($\nu = 0.01$).}
  \label{tab:grid_results_dec_rho}
  \small
  \begin{tabular}{rr|ccc|ccc}
    \toprule
    & & \multicolumn{3}{c|}{$\nu = 0$} & \multicolumn{3}{c}{$\nu = 0.01$} \\
    \cmidrule(lr){3-5} \cmidrule(lr){6-8}
    $n_{\text{pp}}$ & $f_{\text{ov}}$ & Linear & Cubic & Quadratic & Linear & Cubic & Quadratic \\
    \midrule
     10 & 0.01 & 4.97e-01 & 8.04e-05 & 1.21e-04 & 1.64e-02 & 3.25e-04 & 3.63e-04 \\
     15 & 0.01 & 7.56e-03 & \textbf{1.00e-05} & 3.97e-05 & 7.82e-03 & 2.58e-04 & 2.80e-04 \\
     20 & 0.01 & 1.40e-02 & 2.83e-05 & 6.81e-05 & 1.40e-02 & 2.51e-04 & 2.91e-04 \\
     25 & 0.01 & 1.16e-02 & 2.06e-05 & 1.82e-04 & 1.14e-02 & \textbf{2.50e-04} & 4.21e-04 \\
     50 & 0.01 & 8.77e-03 & \textbf{8.33e-06} & 6.72e-04 & 8.83e-03 & \textbf{2.22e-04} & 9.36e-04 \\
    100 & 0.01 & 3.10e-02 & \textbf{1.81e-05} & 1.33e-03 & 3.11e-02 & \textbf{2.14e-04} & 1.63e-03 \\
    \midrule
     10 & 0.05 & 5.66e-03 & 2.51e-05 & 1.13e-04 & 5.90e-03 & 6.07e-04 & 7.07e-04 \\
     15 & 0.05 & 1.01e-02 & 2.96e-03 & 3.52e-03 & 1.08e-02 & 9.16e-04 & 3.52e-03 \\
     20 & 0.05 & 1.47e-02 & 8.61e-03 & 1.33e-03 & 1.45e-02 & 6.63e-03 & 4.08e-03 \\
     25 & 0.05 & 1.91e-02 & 2.76e-05 & 1.73e-04 & 1.92e-02 & 3.08e-04 & 4.46e-04 \\
     50 & 0.05 & 1.38e-02 & 5.09e-05 & 3.75e-03 & 1.47e-02 & 4.40e-04 & 4.70e-03 \\
    100 & 0.05 & 2.10e-02 & 2.07e-04 & 2.27e-03 & 2.14e-02 & 6.59e-04 & 2.53e-03 \\
    \midrule
     10 & 0.10 & 6.47e-03 & 2.29e-04 & 2.31e-04 & 6.82e-03 & 1.41e-03 & 8.61e-04 \\
     15 & 0.10 & 1.07e-02 & 7.51e-04 & 4.79e-03 & 1.11e-02 & 4.15e-04 & 5.32e-03 \\
     20 & 0.10 & 1.19e-02 & 1.47e-03 & 9.32e-03 & 1.20e-02 & 5.50e-03 & 1.60e-02 \\
     25 & 0.10 & 5.73e-02 & 2.99e-04 & 1.22e-03 & 5.55e-02 & 1.29e-03 & 2.17e-03 \\
     50 & 0.10 & 2.02e+00 & 8.56e-05 & 2.44e-03 & 1.99e+00 & 7.10e-04 & 2.99e-03 \\
    100 & 0.10 & 2.59e-02 & 4.52e-04 & 1.43e-03 & 2.60e-02 & 1.10e-03 & 1.87e-03 \\
    \bottomrule
  \end{tabular}
\end{table}

\clearpage


\begin{table}[htbp]
  \centering
  \caption{Held-out test loss by partition size, overlap fraction, and spline
           model (lower is better), at both noise levels. M-surface \texttt{inc\_temp}:
           dataset 3 ($\nu = 0$) and dataset 15 ($\nu = 0.01$).}
  \label{tab:grid_results_inc_temp}
  \small
  \begin{tabular}{rr|ccc|ccc}
    \toprule
    & & \multicolumn{3}{c|}{$\nu = 0$} & \multicolumn{3}{c}{$\nu = 0.01$} \\
    \cmidrule(lr){3-5} \cmidrule(lr){6-8}
    $n_{\text{pp}}$ & $f_{\text{ov}}$ & Linear & Cubic & Quadratic & Linear & Cubic & Quadratic \\
    \midrule
     10 & 0.01 & 2.56e-04 & 2.05e-05 & 1.34e-05 & 4.89e-04 & 2.83e-04 & 2.66e-04 \\
     15 & 0.01 & 3.49e-04 & \textbf{7.15e-06} & 1.24e-05 & 5.75e-04 & 2.37e-04 & 2.40e-04 \\
     20 & 0.01 & 4.44e-04 & \textbf{6.19e-06} & 1.34e-05 & 6.51e-04 & 2.30e-04 & 2.31e-04 \\
     25 & 0.01 & 5.61e-04 & 1.06e-05 & 2.21e-05 & 7.59e-04 & \textbf{2.24e-04} & 2.37e-04 \\
     50 & 0.01 & 1.18e-03 & 8.03e-06 & 5.05e-05 & 1.39e-03 & \textbf{2.14e-04} & 2.46e-04 \\
    100 & 0.01 & 1.86e-03 & \textbf{3.55e-06} & 1.46e-04 & 2.04e-03 & \textbf{2.02e-04} & 3.37e-04 \\
    \midrule
     10 & 0.05 & 2.20e-04 & 4.11e-05 & 2.03e-05 & 4.61e-04 & 3.14e-04 & 2.58e-04 \\
     15 & 0.05 & 3.09e-04 & 5.55e-05 & 4.76e-05 & 5.34e-04 & 3.60e-04 & 2.98e-04 \\
     20 & 0.05 & 4.60e-04 & 2.47e-05 & 4.53e-05 & 6.76e-04 & 5.84e-04 & 3.53e-04 \\
     25 & 0.05 & 5.06e-04 & 1.31e-05 & 2.26e-05 & 7.05e-04 & 2.62e-04 & 2.30e-04 \\
     50 & 0.05 & 1.10e-03 & 4.47e-05 & 5.62e-05 & 1.31e-03 & 3.56e-04 & 2.67e-04 \\
    100 & 0.05 & 1.80e-03 & 2.37e-05 & 1.50e-04 & 1.98e-03 & 2.35e-04 & 3.46e-04 \\
    \midrule
     10 & 0.10 & 1.62e-04 & 1.05e-04 & 4.26e-05 & 3.98e-04 & 2.90e-04 & 2.71e-04 \\
     15 & 0.10 & 2.57e-04 & 2.27e-05 & 3.75e-05 & 4.73e-04 & 2.85e-04 & 2.47e-04 \\
     20 & 0.10 & 4.34e-04 & 3.60e-05 & 1.83e-04 & 6.60e-04 & 4.45e-04 & 4.17e-04 \\
     25 & 0.10 & 2.90e-04 & 3.71e-05 & 4.51e-05 & 4.80e-04 & 3.08e-04 & 2.75e-04 \\
     50 & 0.10 & 9.71e-04 & 6.60e-05 & 1.43e-04 & 1.17e-03 & 5.68e-04 & 2.90e-04 \\
    100 & 0.10 & 1.12e-03 & 1.21e-04 & 1.15e-04 & 1.31e-03 & 2.63e-04 & 3.11e-04 \\
    \bottomrule
  \end{tabular}
\end{table}


\begin{table}[htbp]
  \centering
  \caption{Held-out test loss by partition size, overlap fraction, and spline
           model (lower is better), at both noise levels. M-surface \texttt{dec\_temp}:
           dataset 4 ($\nu = 0$) and dataset 16 ($\nu = 0.01$).}
  \label{tab:grid_results_dec_temp}
  \small
  \begin{tabular}{rr|ccc|ccc}
    \toprule
    & & \multicolumn{3}{c|}{$\nu = 0$} & \multicolumn{3}{c}{$\nu = 0.01$} \\
    \cmidrule(lr){3-5} \cmidrule(lr){6-8}
    $n_{\text{pp}}$ & $f_{\text{ov}}$ & Linear & Cubic & Quadratic & Linear & Cubic & Quadratic \\
    \midrule
     10 & 0.01 & 9.57e-05 & 1.02e-04 & 5.64e-05 & 3.30e-04 & 3.17e-04 & 3.08e-04 \\
     15 & 0.01 & 3.02e-03 & 8.22e-05 & 6.64e-05 & 3.21e-03 & 3.03e-04 & 3.06e-04 \\
     20 & 0.01 & 2.08e-04 & 2.26e-04 & 2.11e-04 & 4.28e-04 & 4.20e-04 & 4.30e-04 \\
     25 & 0.01 & 2.74e-04 & 1.88e-04 & 1.87e-04 & 4.91e-04 & 3.75e-04 & 4.40e-04 \\
     50 & 0.01 & 2.49e-04 & 1.63e-04 & 1.75e-04 & 4.50e-04 & 4.03e-04 & 3.54e-04 \\
    100 & 0.01 & 9.54e-05 & \textbf{1.45e-05} & 2.82e-05 & 2.68e-04 & \textbf{2.12e-04} & 2.22e-04 \\
    \midrule
     10 & 0.05 & 7.52e-04 & 6.41e-05 & 6.31e-05 & 1.08e-03 & 4.28e-04 & 3.04e-04 \\
     15 & 0.05 & 1.34e-04 & 8.92e-05 & 8.66e-05 & 3.38e-04 & 3.75e-04 & 2.92e-04 \\
     20 & 0.05 & 7.18e-04 & 3.85e-05 & 3.85e-05 & 4.07e-04 & 3.26e-04 & 2.82e-04 \\
     25 & 0.05 & 4.84e-04 & 2.06e-04 & 1.92e-04 & 6.93e-04 & 3.74e-04 & 5.87e-04 \\
     50 & 0.05 & 2.22e-04 & 6.63e-05 & 3.27e-05 & 4.14e-04 & 2.69e-04 & 2.34e-04 \\
    100 & 0.05 & 6.72e-05 & \textbf{1.34e-05} & 2.10e-05 & 2.71e-04 & \textbf{2.08e-04} & \textbf{2.21e-04} \\
    \midrule
     10 & 0.10 & 7.61e-04 & 7.56e-05 & 7.80e-05 & 7.47e-04 & 4.04e-04 & 5.45e-04 \\
     15 & 0.10 & 1.49e-04 & 2.38e-04 & 1.51e-04 & 3.27e-04 & 4.78e-04 & 3.03e-04 \\
     20 & 0.10 & 2.82e-04 & 5.87e-05 & 5.75e-05 & 5.04e-04 & 3.60e-04 & 3.12e-04 \\
     25 & 0.10 & 9.94e-05 & 9.05e-05 & 9.68e-05 & 3.06e-04 & 3.59e-04 & 3.24e-04 \\
     50 & 0.10 & 5.33e-05 & 2.12e-05 & \textbf{1.32e-05} & 3.91e-04 & 2.33e-04 & 2.36e-03 \\
    100 & 0.10 & 5.43e-05 & 2.81e-05 & 3.76e-05 & 2.50e-04 & 2.51e-04 & 2.30e-04 \\
    \bottomrule
  \end{tabular}
\end{table}

\clearpage


\begin{table}[htbp]
  \centering
  \caption{Held-out test loss by partition size, overlap fraction, and spline
           model (lower is better), at both noise levels. M-surface \texttt{inc\_rho\_temp}:
           dataset 5 ($\nu = 0$) and dataset 17 ($\nu = 0.01$).}
  \label{tab:grid_results_inc_rho_temp}
  \small
  \begin{tabular}{rr|ccc|ccc}
    \toprule
    & & \multicolumn{3}{c|}{$\nu = 0$} & \multicolumn{3}{c}{$\nu = 0.01$} \\
    \cmidrule(lr){3-5} \cmidrule(lr){6-8}
    $n_{\text{pp}}$ & $f_{\text{ov}}$ & Linear & Cubic & Quadratic & Linear & Cubic & Quadratic \\
    \midrule
     10 & 0.01 & 5.27e-05 & \textbf{5.08e-06} & \textbf{5.39e-06} & 3.06e-04 & 2.68e-04 & 2.65e-04 \\
     15 & 0.01 & 8.25e-05 & 5.83e-06 & 7.52e-06 & 2.90e-04 & 2.32e-04 & 2.31e-04 \\
     20 & 0.01 & 1.01e-04 & 6.91e-06 & 8.85e-06 & 3.06e-04 & 2.17e-04 & 2.17e-04 \\
     25 & 0.01 & 1.32e-04 & 9.10e-06 & 1.28e-05 & 3.29e-04 & 2.13e-04 & 2.16e-04 \\
     50 & 0.01 & 2.76e-04 & 8.06e-06 & 4.06e-05 & 4.58e-04 & \textbf{2.09e-04} & 2.33e-04 \\
    100 & 0.01 & 4.87e-04 & \textbf{5.15e-06} & 8.28e-05 & 6.61e-04 & \textbf{1.96e-04} & 2.71e-04 \\
    \midrule
     10 & 0.05 & 4.70e-05 & 2.22e-05 & 2.08e-05 & 3.20e-04 & 2.65e-04 & 2.52e-04 \\
     15 & 0.05 & 6.97e-05 & 1.11e-05 & 1.27e-05 & 2.99e-04 & 2.30e-04 & 2.35e-04 \\
     20 & 0.05 & 9.23e-05 & 1.72e-05 & 1.25e-05 & 3.04e-04 & 2.21e-04 & 2.22e-04 \\
     25 & 0.05 & 1.06e-04 & 2.26e-05 & 1.61e-05 & 3.05e-04 & 2.32e-04 & 2.31e-04 \\
     50 & 0.05 & 2.27e-04 & 2.62e-05 & 8.45e-05 & 4.28e-04 & 2.63e-04 & 2.72e-04 \\
    100 & 0.05 & 4.41e-04 & 8.18e-06 & 8.03e-05 & 6.19e-04 & \textbf{1.95e-04} & 4.29e-04 \\
    \midrule
     10 & 0.10 & 3.62e-05 & 4.27e-05 & 3.52e-05 & 5.87e-04 & 2.71e-04 & 2.50e-04 \\
     15 & 0.10 & 4.78e-05 & 1.21e-05 & 1.57e-05 & 2.58e-04 & 6.74e-04 & 9.51e-04 \\
     20 & 0.10 & 6.79e-05 & 1.56e-05 & 2.51e-05 & 2.82e-04 & 2.60e-04 & 3.54e-04 \\
     25 & 0.10 & 7.19e-05 & 1.12e-05 & 4.05e-05 & 2.75e-04 & 2.34e-04 & 2.18e-04 \\
     50 & 0.10 & 2.00e-04 & 2.26e-05 & 9.20e-05 & 4.16e-04 & 2.80e-04 & 2.91e-04 \\
    100 & 0.10 & 4.33e-04 & 3.15e-05 & 1.82e-04 & 6.40e-04 & 2.44e-04 & 3.82e-04 \\
    \bottomrule
  \end{tabular}
\end{table}


\begin{table}[htbp]
  \centering
  \caption{Held-out test loss by partition size, overlap fraction, and spline
           model (lower is better), at both noise levels. M-surface \texttt{dec\_rho\_temp}:
           dataset 6 ($\nu = 0$) and dataset 18 ($\nu = 0.01$).}
  \label{tab:grid_results_dec_rho_temp}
  \small
  \begin{tabular}{rr|ccc|ccc}
    \toprule
    & & \multicolumn{3}{c|}{$\nu = 0$} & \multicolumn{3}{c}{$\nu = 0.01$} \\
    \cmidrule(lr){3-5} \cmidrule(lr){6-8}
    $n_{\text{pp}}$ & $f_{\text{ov}}$ & Linear & Cubic & Quadratic & Linear & Cubic & Quadratic \\
    \midrule
     10 & 0.01 & 9.78e-01 & 4.39e+02 & 2.17e-01 & 1.02e+00 & 1.15e+00 & 1.31e+00 \\
     15 & 0.01 & 9.15e-01 & 2.37e-02 & 2.89e-02 & 9.09e-01 & \textbf{2.03e-02} & 2.94e-02 \\
     20 & 0.01 & 1.38e-01 & 3.19e-02 & 2.91e-02 & 1.40e-01 & 2.50e-02 & 2.80e-02 \\
     25 & 0.01 & 1.25e+00 & 2.31e-02 & 5.97e-02 & 1.28e+00 & 2.49e-02 & 5.18e-02 \\
     50 & 0.01 & 4.98e-01 & 3.86e-02 & 4.76e-02 & 4.90e-01 & 3.80e-02 & 4.76e-02 \\
    100 & 0.01 & 2.72e+00 & \textbf{2.40e-03} & 2.24e-02 & 2.78e+00 & \textbf{2.70e-03} & \textbf{1.06e-02} \\
    \midrule
     10 & 0.05 & 3.21e+00 & 3.28e+02 & 2.71e-02 & 3.34e+00 & 1.80e-01 & 1.95e-01 \\
     15 & 0.05 & 9.79e-01 & 2.28e-02 & 3.40e-02 & 9.67e-01 & 3.80e-02 & 3.24e-02 \\
     20 & 0.05 & 2.18e-01 & 5.73e-02 & 6.50e-02 & 1.79e-01 & 5.40e-02 & 7.50e-02 \\
     25 & 0.05 & 1.13e+00 & 2.87e-02 & 5.23e-02 & 1.18e+00 & 2.88e-02 & 5.61e-02 \\
     50 & 0.05 & 1.15e+00 & 3.31e-02 & 4.35e-02 & 7.74e-01 & 3.25e-02 & 4.18e-02 \\
    100 & 0.05 & 6.21e-01 & \textbf{3.46e-03} & 4.90e-02 & 6.43e-01 & 2.78e-02 & 3.76e-02 \\
    \midrule
     10 & 0.10 & 1.73e+01 & 5.76e+05 & 3.34e+04 & 1.17e+02 & 2.11e+05 & 4.15e+05 \\
     15 & 0.10 & 2.18e+00 & 4.86e-02 & 5.35e-02 & 1.66e+00 & 2.82e-02 & 5.34e-02 \\
     20 & 0.10 & 1.17e-01 & 2.23e-02 & 3.12e-02 & 1.19e-01 & 2.64e-02 & 2.45e-02 \\
     25 & 0.10 & 1.00e-01 & 2.05e-02 & 2.33e-02 & 1.04e-01 & 2.11e-02 & 2.44e-02 \\
     50 & 0.10 & 4.26e-01 & 3.19e-02 & 2.74e-02 & 4.35e-01 & 3.25e-02 & 3.09e-02 \\
    100 & 0.10 & 8.90e-01 & \textbf{5.85e-03} & 6.28e-02 & 8.68e-01 & 3.40e-02 & 5.54e-02 \\
    \bottomrule
  \end{tabular}
\end{table}

\clearpage


\begin{table}[htbp]
  \centering
  \caption{Held-out test loss by partition size, overlap fraction, and spline
           model (lower is better), at both noise levels. M-surface \texttt{inc\_temp\_dec\_rho}:
           dataset 7 ($\nu = 0$) and dataset 19 ($\nu = 0.01$).}
  \label{tab:grid_results_inc_temp_dec_rho}
  \small
  \begin{tabular}{rr|ccc|ccc}
    \toprule
    & & \multicolumn{3}{c|}{$\nu = 0$} & \multicolumn{3}{c}{$\nu = 0.01$} \\
    \cmidrule(lr){3-5} \cmidrule(lr){6-8}
    $n_{\text{pp}}$ & $f_{\text{ov}}$ & Linear & Cubic & Quadratic & Linear & Cubic & Quadratic \\
    \midrule
     10 & 0.01 & 6.62e-02 & 1.11e-03 & 1.50e-03 & 9.54e-02 & 1.01e-03 & 1.26e-03 \\
     15 & 0.01 & 2.89e-02 & \textbf{1.25e-04} & \textbf{4.10e-04} & 3.43e-02 & \textbf{5.38e-04} & \textbf{7.49e-04} \\
     20 & 0.01 & 1.37e-01 & 4.16e-04 & 7.01e-04 & 1.41e-01 & 9.13e-04 & 9.81e-04 \\
     25 & 0.01 & 1.78e-01 & 1.12e-03 & 1.61e-03 & 1.94e-01 & 1.26e-03 & 1.71e-03 \\
     50 & 0.01 & 2.18e-01 & 1.35e-03 & 4.74e-03 & 2.21e-01 & 1.03e-03 & 3.53e-03 \\
    100 & 0.01 & 2.32e-01 & \textbf{4.16e-04} & 5.20e-03 & 2.33e-01 & \textbf{7.02e-04} & 4.97e-03 \\
    \midrule
     10 & 0.05 & 1.11e-01 & 6.41e-03 & 7.93e-03 & 1.22e-01 & 1.57e-01 & 4.48e-02 \\
     15 & 0.05 & 1.02e-01 & 4.58e-02 & 5.84e-02 & 9.85e-02 & 8.14e-02 & 4.36e-02 \\
     20 & 0.05 & 1.76e-01 & 5.29e-03 & 1.04e-03 & 1.79e-01 & 1.85e-02 & 2.46e-03 \\
     25 & 0.05 & 1.93e-01 & 2.20e-03 & 2.30e-03 & 2.03e-01 & 2.63e-03 & 2.56e-03 \\
     50 & 0.05 & 3.33e-01 & 5.30e-03 & 2.56e-02 & 3.37e-01 & 1.08e-02 & 9.12e-03 \\
    100 & 0.05 & 3.70e-01 & 4.99e-03 & 7.93e-03 & 3.79e-01 & 7.02e-03 & 8.99e-03 \\
    \midrule
     10 & 0.10 & 1.79e+00 & 2.78e-02 & 2.52e-02 & 1.64e+00 & 7.60e-02 & 1.11e-01 \\
     15 & 0.10 & 1.64e-01 & 7.61e-02 & 5.75e-02 & 1.75e-01 & 1.12e-01 & 4.32e-02 \\
     20 & 0.10 & 2.49e-01 & 1.66e-02 & 3.52e-02 & 2.56e-01 & 4.76e-02 & 4.90e-02 \\
     25 & 0.10 & 1.95e+00 & 1.68e-01 & 1.12e-02 & 1.90e+00 & 1.48e-01 & 2.44e-02 \\
     50 & 0.10 & 1.26e+00 & 1.14e-02 & 9.59e-03 & 9.60e-01 & 2.26e-02 & 4.18e-02 \\
    100 & 0.10 & 4.83e-01 & 7.07e-03 & 1.50e-02 & 4.87e-01 & 1.57e-02 & 1.34e-02 \\
    \bottomrule
  \end{tabular}
\end{table}


\begin{table}[htbp]
  \centering
  \caption{Held-out test loss by partition size, overlap fraction, and spline
           model (lower is better), at both noise levels. M-surface \texttt{inc\_rho\_dec\_temp}:
           dataset 8 ($\nu = 0$) and dataset 20 ($\nu = 0.01$).}
  \label{tab:grid_results_inc_rho_dec_temp}
  \small
  \begin{tabular}{rr|ccc|ccc}
    \toprule
    & & \multicolumn{3}{c|}{$\nu = 0$} & \multicolumn{3}{c}{$\nu = 0.01$} \\
    \cmidrule(lr){3-5} \cmidrule(lr){6-8}
    $n_{\text{pp}}$ & $f_{\text{ov}}$ & Linear & Cubic & Quadratic & Linear & Cubic & Quadratic \\
    \midrule
     10 & 0.01 & 2.54e-04 & 2.06e-04 & 2.05e-04 & 5.19e-04 & 4.73e-04 & 4.72e-04 \\
     15 & 0.01 & 3.53e-04 & 3.46e-04 & 8.35e-04 & 8.46e-04 & 6.20e-04 & 6.39e-04 \\
     20 & 0.01 & 3.21e-04 & 2.27e-04 & 2.33e-04 & 5.41e-04 & 4.58e-04 & 4.57e-04 \\
     25 & 0.01 & 7.26e-04 & 6.21e-04 & 6.17e-04 & 9.20e-04 & 8.36e-04 & 7.99e-04 \\
     50 & 0.01 & 5.53e-04 & 8.86e-04 & 4.07e-04 & 7.65e-04 & 6.01e-04 & 6.05e-04 \\
    100 & 0.01 & 2.21e-04 & 8.66e-05 & 1.01e-04 & 4.25e-04 & 2.94e-04 & 3.03e-04 \\
    \midrule
     10 & 0.05 & 2.54e-04 & 2.00e-04 & 2.00e-04 & 5.80e-04 & 5.47e-04 & 4.58e-04 \\
     15 & 0.05 & 3.07e-04 & 3.63e-04 & 8.82e-04 & 8.78e-04 & 7.16e-04 & 1.15e-03 \\
     20 & 0.05 & 3.33e-04 & 2.50e-04 & 2.25e-04 & 5.69e-04 & 6.70e-04 & 5.30e-04 \\
     25 & 0.05 & 2.83e-03 & 4.82e-04 & 4.01e-04 & 6.73e-04 & 6.62e-04 & 5.33e-04 \\
     50 & 0.05 & 4.45e-04 & 2.87e-04 & 2.99e-04 & 6.82e-04 & 4.92e-04 & 4.65e-04 \\
    100 & 0.05 & \textbf{4.98e-05} & 7.61e-05 & 5.66e-05 & \textbf{2.27e-04} & 3.31e-04 & 2.69e-04 \\
    \midrule
     10 & 0.10 & 2.54e-04 & 2.14e-04 & 2.15e-04 & 5.18e-04 & 7.10e-04 & 5.48e-04 \\
     15 & 0.10 & 2.60e-04 & 3.21e-04 & 2.08e-04 & 6.18e-04 & 7.21e-04 & 5.32e-04 \\
     20 & 0.10 & 3.36e-04 & 3.31e-04 & 1.90e-04 & 6.99e-04 & 7.50e-04 & 6.26e-04 \\
     25 & 0.10 & 3.16e-04 & 2.93e-04 & 2.40e-04 & 5.61e-04 & 7.09e-04 & 3.83e-04 \\
     50 & 0.10 & 3.12e-04 & 3.32e-04 & 2.55e-04 & 6.18e-04 & 5.98e-04 & 5.06e-04 \\
    100 & 0.10 & \textbf{4.15e-05} & 9.87e-05 & \textbf{5.36e-05} & \textbf{2.23e-04} & 3.38e-04 & \textbf{2.59e-04} \\
    \bottomrule
  \end{tabular}
\end{table}

\clearpage


\begin{table}[htbp]
  \centering
  \caption{Held-out test loss by partition size, overlap fraction, and spline
           model (lower is better), at both noise levels. M-surface \texttt{parabola\_rho}:
           dataset 9 ($\nu = 0$) and dataset 21 ($\nu = 0.01$).}
  \label{tab:grid_results_parabola_rho}
  \small
  \begin{tabular}{rr|ccc|ccc}
    \toprule
    & & \multicolumn{3}{c|}{$\nu = 0$} & \multicolumn{3}{c}{$\nu = 0.01$} \\
    \cmidrule(lr){3-5} \cmidrule(lr){6-8}
    $n_{\text{pp}}$ & $f_{\text{ov}}$ & Linear & Cubic & Quadratic & Linear & Cubic & Quadratic \\
    \midrule
     10 & 0.01 & 5.35e-04 & 1.40e-04 & 1.22e-04 & 9.63e-04 & 4.01e-04 & 3.87e-04 \\
     15 & 0.01 & 2.71e-03 & 1.23e-03 & 8.69e-04 & 2.82e-03 & 4.02e-04 & 3.95e-04 \\
     20 & 0.01 & 4.98e-04 & 2.06e-04 & 1.07e-04 & 7.20e-04 & 3.53e-04 & \textbf{3.46e-04} \\
     25 & 0.01 & 1.17e-03 & 1.56e-04 & 2.23e-04 & 1.33e-03 & 7.00e-04 & 4.97e-04 \\
     50 & 0.01 & 2.93e-03 & \textbf{9.68e-05} & 9.80e-05 & 3.49e-03 & 3.62e-04 & 3.81e-04 \\
    100 & 0.01 & 2.38e-03 & \textbf{8.23e-05} & \textbf{8.66e-05} & 2.57e-03 & \textbf{2.73e-04} & \textbf{2.96e-04} \\
    \midrule
     10 & 0.05 & 8.84e-04 & 6.12e-04 & 5.08e-04 & 1.03e-03 & 1.24e-03 & 1.21e-03 \\
     15 & 0.05 & 4.86e-04 & 1.28e-03 & 8.78e-04 & 6.75e-04 & 7.96e-04 & 7.02e-04 \\
     20 & 0.05 & 7.53e-04 & 1.23e-03 & 1.54e-03 & 9.60e-04 & 1.76e-03 & 2.46e-03 \\
     25 & 0.05 & 1.99e-03 & 2.64e-04 & 2.73e-04 & 1.70e-03 & 1.04e-03 & 8.00e-04 \\
     50 & 0.05 & 1.76e-02 & 8.14e-04 & 6.51e-04 & 1.93e-02 & 6.73e-04 & 1.71e-03 \\
    100 & 0.05 & 2.29e-03 & 5.69e-04 & 6.40e-04 & 2.49e-03 & 1.14e-03 & 1.31e-03 \\
    \midrule
     10 & 0.10 & 4.30e-03 & 4.07e-03 & 4.52e-03 & 5.27e-03 & 6.09e-03 & 5.00e-03 \\
     15 & 0.10 & 5.90e-04 & 2.49e-03 & 9.96e-04 & 6.68e-04 & 1.43e-03 & 1.54e-03 \\
     20 & 0.10 & 1.29e-03 & 1.52e-03 & 8.22e-04 & 1.46e-03 & 2.06e-03 & 2.81e-03 \\
     25 & 0.10 & 2.77e-02 & 2.76e-03 & 3.23e-03 & 3.06e-02 & 9.39e-03 & 4.66e-03 \\
     50 & 0.10 & 1.34e-02 & 2.11e-02 & 9.50e-03 & 2.16e-02 & 3.34e-02 & 1.50e-02 \\
    100 & 0.10 & 4.57e-03 & 1.75e-02 & 1.25e-03 & 5.10e-03 & 1.38e-02 & 1.30e-03 \\
    \bottomrule
  \end{tabular}
\end{table}


\begin{table}[htbp]
  \centering
  \caption{Held-out test loss by partition size, overlap fraction, and spline
           model (lower is better), at both noise levels. M-surface \texttt{parabola\_temp}:
           dataset 10 ($\nu = 0$) and dataset 22 ($\nu = 0.01$).}
  \label{tab:grid_results_parabola_temp}
  \small
  \begin{tabular}{rr|ccc|ccc}
    \toprule
    & & \multicolumn{3}{c|}{$\nu = 0$} & \multicolumn{3}{c}{$\nu = 0.01$} \\
    \cmidrule(lr){3-5} \cmidrule(lr){6-8}
    $n_{\text{pp}}$ & $f_{\text{ov}}$ & Linear & Cubic & Quadratic & Linear & Cubic & Quadratic \\
    \midrule
     10 & 0.01 & 1.66e-03 & 7.07e-04 & 4.03e-04 & 2.60e-03 & 8.72e-03 & 4.75e-03 \\
     15 & 0.01 & 1.32e-03 & \textbf{1.40e-05} & \textbf{1.37e-05} & 1.56e-03 & 2.48e-04 & 2.50e-04 \\
     20 & 0.01 & 9.63e-04 & 1.46e-05 & 1.58e-05 & 1.18e-03 & \textbf{2.34e-04} & 2.41e-04 \\
     25 & 0.01 & 2.23e-03 & 3.12e-05 & 3.90e-05 & 2.44e-03 & 2.42e-04 & 2.40e-04 \\
     50 & 0.01 & 7.56e-03 & 1.07e-04 & 1.38e-04 & 7.80e-03 & 3.05e-04 & 3.24e-04 \\
    100 & 0.01 & 3.08e-03 & 3.10e-05 & 5.76e-05 & 3.29e-03 & \textbf{2.29e-04} & 2.52e-04 \\
    \midrule
     10 & 0.05 & 1.51e-03 & 1.33e-03 & 4.12e-04 & 1.68e-03 & 9.70e-03 & 6.52e-03 \\
     15 & 0.05 & 1.32e-03 & 1.11e-03 & 1.17e-03 & 1.55e-03 & 4.98e-03 & 3.69e-03 \\
     20 & 0.05 & 9.95e-04 & \textbf{1.28e-05} & 2.52e-05 & 2.48e-03 & \textbf{2.35e-04} & 2.38e-04 \\
     25 & 0.05 & 2.52e-03 & 2.08e-05 & 3.18e-05 & 2.50e-03 & 2.44e-04 & 2.52e-04 \\
     50 & 0.05 & 4.35e-03 & 3.40e-05 & 5.53e-05 & 4.58e-03 & 4.62e-04 & 4.98e-04 \\
    100 & 0.05 & 3.64e-03 & 3.74e-03 & 3.68e-02 & 3.86e-03 & 4.30e-02 & 6.30e-02 \\
    \midrule
     10 & 0.10 & 2.77e-03 & 3.76e-03 & 3.33e-03 & 3.07e-03 & 2.81e-02 & 4.37e-02 \\
     15 & 0.10 & 5.98e-04 & 1.09e-03 & 8.48e-04 & 8.28e-04 & 8.32e-03 & 8.80e-03 \\
     20 & 0.10 & 8.31e-04 & 9.68e-04 & 1.06e-03 & 1.03e-03 & 2.07e-02 & 3.58e-03 \\
     25 & 0.10 & 2.76e-03 & 2.01e-04 & 4.45e-05 & 2.88e-03 & 2.29e-03 & 2.58e-03 \\
     50 & 0.10 & 3.87e-03 & 2.77e-03 & 5.88e-03 & 4.01e-03 & 1.51e-02 & 8.77e-03 \\
    100 & 0.10 & 4.05e-03 & 2.81e-03 & 2.98e-02 & 4.25e-03 & 3.08e-03 & 4.99e-03 \\
    \bottomrule
  \end{tabular}
\end{table}

\clearpage


\begin{table}[htbp]
  \centering
  \caption{Held-out test loss by partition size, overlap fraction, and spline
           model (lower is better), at both noise levels. M-surface \texttt{sin\_basis}:
           dataset 11 ($\nu = 0$) and dataset 23 ($\nu = 0.01$).}
  \label{tab:grid_results_sin_basis}
  \small
  \begin{tabular}{rr|ccc|ccc}
    \toprule
    & & \multicolumn{3}{c|}{$\nu = 0$} & \multicolumn{3}{c}{$\nu = 0.01$} \\
    \cmidrule(lr){3-5} \cmidrule(lr){6-8}
    $n_{\text{pp}}$ & $f_{\text{ov}}$ & Linear & Cubic & Quadratic & Linear & Cubic & Quadratic \\
    \midrule
     10 & 0.01 & 6.97e-03 & \textbf{1.31e-04} & \textbf{1.83e-04} & 7.32e-03 & 4.72e-04 & 4.63e-04 \\
     15 & 0.01 & 1.34e-02 & 2.51e-04 & 3.60e-04 & 9.83e-03 & \textbf{3.44e-04} & \textbf{3.60e-04} \\
     20 & 0.01 & 8.70e-03 & 1.92e-04 & 5.88e-04 & 1.00e-02 & 5.07e-04 & 8.60e-04 \\
     25 & 0.01 & 1.36e-02 & 3.47e-04 & 3.82e-04 & 1.38e-02 & 7.91e-04 & 7.37e-04 \\
     50 & 0.01 & 2.20e-02 & \textbf{1.52e-04} & 2.52e-04 & 2.25e-02 & \textbf{3.72e-04} & 5.90e-04 \\
    100 & 0.01 & 5.58e-02 & 3.47e-04 & 6.80e-04 & 5.61e-02 & 4.74e-04 & 9.22e-04 \\
    \midrule
     10 & 0.05 & 2.52e-02 & 2.52e-04 & 2.19e-04 & 2.49e-02 & 9.97e-04 & 6.50e-04 \\
     15 & 0.05 & 1.58e-02 & 1.54e-03 & 1.78e-03 & 1.44e-02 & 5.34e-03 & 3.64e-03 \\
     20 & 0.05 & 3.13e-01 & 3.44e-02 & 7.58e-03 & 2.24e-02 & 4.91e-02 & 3.30e-02 \\
     25 & 0.05 & 1.36e-02 & 8.68e-03 & 1.96e-03 & 1.40e-02 & 1.40e-02 & 1.45e-02 \\
     50 & 0.05 & 7.60e-02 & 7.41e-04 & 7.62e-04 & 7.81e-02 & 1.09e-03 & 1.07e-03 \\
    100 & 0.05 & 9.79e-02 & 1.13e-03 & 1.62e-03 & 9.83e-02 & 1.70e-03 & 1.97e-03 \\
    \midrule
     10 & 0.10 & 6.19e+00 & 5.00e-04 & 2.81e-04 & 2.25e-02 & 1.52e-03 & 1.11e-03 \\
     15 & 0.10 & 3.70e-02 & 9.29e-04 & 1.24e-03 & 3.80e-02 & 3.72e-03 & 3.28e-03 \\
     20 & 0.10 & 2.06e-02 & 2.21e-02 & 1.73e-02 & 2.51e-02 & 3.88e-02 & 2.86e-02 \\
     25 & 0.10 & 3.57e-02 & 1.69e-03 & 9.83e-04 & 3.70e-02 & 3.76e-03 & 2.91e-03 \\
     50 & 0.10 & 8.41e-02 & 6.89e-04 & 6.54e-04 & 7.96e-02 & 1.74e-03 & 1.72e-03 \\
    100 & 0.10 & 8.74e-02 & 2.65e-03 & 2.18e-03 & 8.81e-02 & 3.70e-03 & 2.60e-03 \\
    \bottomrule
  \end{tabular}
\end{table}


\begin{table}[htbp]
  \centering
  \caption{Held-out test loss by partition size, overlap fraction, and spline
           model (lower is better), at both noise levels. M-surface \texttt{mixed\_basis}:
           dataset 12 ($\nu = 0$) and dataset 24 ($\nu = 0.01$).}
  \label{tab:grid_results_mixed_basis}
  \small
  \begin{tabular}{rr|ccc|ccc}
    \toprule
    & & \multicolumn{3}{c|}{$\nu = 0$} & \multicolumn{3}{c}{$\nu = 0.01$} \\
    \cmidrule(lr){3-5} \cmidrule(lr){6-8}
    $n_{\text{pp}}$ & $f_{\text{ov}}$ & Linear & Cubic & Quadratic & Linear & Cubic & Quadratic \\
    \midrule
     10 & 0.01 & 1.18e-01 & \textbf{2.04e-04} & \textbf{1.72e-04} & 1.19e-01 & 4.60e-04 & 5.28e-04 \\
     15 & 0.01 & 1.15e-01 & 2.50e-04 & \textbf{1.96e-04} & 1.16e-01 & \textbf{3.78e-04} & 4.17e-04 \\
     20 & 0.01 & 2.05e-01 & 2.62e-04 & 3.09e-04 & 2.14e-01 & \textbf{3.97e-04} & 4.70e-04 \\
     25 & 0.01 & 3.21e-01 & 4.81e-04 & 5.34e-04 & 3.21e-01 & 4.79e-04 & 8.20e-04 \\
     50 & 0.01 & 2.35e-01 & 2.60e-04 & 1.15e-03 & 2.38e-01 & \textbf{4.03e-04} & 1.39e-03 \\
    100 & 0.01 & 3.33e-01 & 7.50e-04 & 2.58e-03 & 3.30e-01 & 6.62e-04 & 2.82e-03 \\
    \midrule
     10 & 0.05 & 2.50e-01 & 2.52e-04 & 1.23e-03 & 1.83e-01 & 6.37e-04 & 2.44e-03 \\
     15 & 0.05 & 5.31e-01 & 2.74e-03 & 2.77e-03 & 8.87e-01 & 2.40e-03 & 3.25e-03 \\
     20 & 0.05 & 5.34e-01 & 3.72e-04 & 3.97e-04 & 5.41e-01 & 6.71e-04 & 7.41e-04 \\
     25 & 0.05 & 4.77e-01 & 5.22e-04 & 8.45e-04 & 2.01e+00 & 7.53e-04 & 2.12e-03 \\
     50 & 0.05 & 6.06e+00 & 3.94e-04 & 1.38e-03 & 3.52e+00 & 1.00e-03 & 1.60e-03 \\
    100 & 0.05 & 3.88e-01 & 9.51e-04 & 3.67e-03 & 3.89e-01 & 7.44e-04 & 3.79e-03 \\
    \midrule
     10 & 0.10 & 1.47e+00 & 1.13e-03 & 8.70e-04 & 3.40e-01 & 7.16e-04 & 9.83e-04 \\
     15 & 0.10 & 1.20e+00 & 7.86e-04 & 9.23e-04 & 5.29e+00 & 1.77e-03 & 4.96e-03 \\
     20 & 0.10 & 1.58e+00 & 6.86e-04 & 1.51e-03 & 1.66e+00 & 7.76e-04 & 1.83e-03 \\
     25 & 0.10 & 3.35e+00 & 8.12e-04 & 1.04e-03 & 3.38e+00 & 9.15e-04 & 1.57e-03 \\
     50 & 0.10 & 3.45e+00 & 1.50e-03 & 2.25e-03 & 1.28e+00 & 1.22e-03 & 2.44e-03 \\
    100 & 0.10 & 7.39e-01 & 1.21e-03 & 3.95e-03 & 6.22e-01 & 1.00e-03 & 4.22e-03 \\
    \bottomrule
  \end{tabular}
\end{table}

%% file: global_rbf_simulated_error_table.tex
\begin{table}[ht]
\centering
\caption{Held-out relative error and test loss by case and basis size $k$, noise level $\nu = 0$.}
\label{tab:results_nu0}
\begin{tabular}{r|r|cc|c}
  \toprule
  Case & $k$ & Eion Rel.\ Error & Pion Rel.\ Error & Test Rel.\ Loss \\
  \midrule
   1 & 16 & 0.012881 & 0.014382 & 0.027262 \\
   1 & 25 & 0.014378 & 0.015863 & 0.030241 \\
  \hline
   2 & 16 & 0.017388 & 0.099033 & 0.116422 \\
   2 & 25 & 0.025470 & 0.114352 & 0.139822 \\
  \hline
   3 & 16 & 0.230539 & 0.030891 & 0.261430 \\
   3 & 25 & 0.256076 & 0.028507 & 0.284583 \\
  \hline
   4 & 16 & 0.028542 & 0.006591 & 0.035133 \\
   4 & 25 & 0.022706 & 0.010807 & 0.033513 \\
  \hline
   5 & 16 & 0.245158 & 0.131288 & 0.787951 \\
   5 & 25 & 0.272318 & 0.118979 & 0.736777 \\
  \hline
   6 & 16 & 0.026039 & 0.277778 & 0.303817 \\
   6 & 25 & 0.027901 & 0.239975 & 0.267876 \\
  \hline
   7 & 16 & 1.295643 & 2.831114 & 4.126757 \\
   7 & 25 & 0.789849 & 1.838996 & 2.628844 \\
  \hline
   8 & 16 & 0.005941 & 0.007305 & 0.013246 \\
   8 & 25 & 0.005738 & 0.007058 & 0.012796 \\
  \hline
   9 & 16 & 0.026660 & 0.108718 & 0.135378 \\
   9 & 25 & 0.025104 & 0.077179 & 0.102283 \\
  \hline
  10 & 16 & 0.185639 & 0.020828 & 0.206467 \\
  10 & 25 & 0.119727 & 0.025883 & 0.145610 \\
  \hline
  11 & 16 & 0.053721 & 0.192916 & 0.246637 \\
  11 & 25 & 0.038200 & 0.096651 & 0.134850 \\
  \hline
  12 & 16 & 0.417087 & 0.683957 & 1.101044 \\
  12 & 25 & 0.439015 & 0.553749 & 0.992764 \\
  \bottomrule
\end{tabular}
\end{table}

\begin{table}[ht]
\centering
\caption{Held-out relative error and test loss by case and basis size $k$, noise level $\nu = 0.01$.}
\label{tab:results_nu01}
\begin{tabular}{r|r|cc|c}
  \toprule
  Case & $k$ & Eion Rel.\ Error & Pion Rel.\ Error & Test Rel.\ Loss \\
  \midrule
   1 & 16 & 0.012940 & 0.014460 & 0.027400 \\
   1 & 25 & 0.014422 & 0.015956 & 0.030378 \\
  \hline
   2 & 16 & 0.017481 & 0.099263 & 0.116743 \\
   2 & 25 & 0.025574 & 0.114536 & 0.140110 \\
  \hline
   3 & 16 & 0.230155 & 0.030840 & 0.260995 \\
   3 & 25 & 0.255688 & 0.028552 & 0.284241 \\
  \hline
   4 & 16 & 0.028597 & 0.006711 & 0.035309 \\
   4 & 25 & 0.022786 & 0.010949 & 0.033735 \\
  \hline
   5 & 16 & 0.245415 & 0.132037 & 0.790087 \\
   5 & 25 & 0.272066 & 0.119708 & 0.738860 \\
  \hline
   6 & 16 & 0.025938 & 0.275659 & 0.301597 \\
   6 & 25 & 0.027844 & 0.240983 & 0.268827 \\
  \hline
   7 & 16 & 1.295670 & 2.823413 & 4.119084 \\
   7 & 25 & 0.790103 & 1.837210 & 2.627313 \\
  \hline
   8 & 16 & 0.006028 & 0.007418 & 0.013447 \\
   8 & 25 & 0.005826 & 0.007173 & 0.012999 \\
  \hline
   9 & 16 & 0.026708 & 0.108800 & 0.135509 \\
   9 & 25 & 0.025230 & 0.077256 & 0.102486 \\
  \hline
  10 & 16 & 0.185656 & 0.020724 & 0.206381 \\
  10 & 25 & 0.119707 & 0.025804 & 0.145512 \\
  \hline
  11 & 16 & 0.053732 & 0.192973 & 0.246705 \\
  11 & 25 & 0.038223 & 0.096682 & 0.134905 \\
  \hline
  12 & 16 & 0.417255 & 0.683995 & 1.101250 \\
  12 & 25 & 0.438630 & 0.554573 & 0.993203 \\
  \bottomrule
\end{tabular}
\end{table}

%% file: main.bbl
\begin{thebibliography}{}

\bibitem[Ali et~al., 2020]{ali2020development}
Ali, S., Swift, D., Wu, C., and Kraus, R. (2020).
\newblock Development of uncertainty-aware equation-of-state models:
  Application to copper.
\newblock {\em Journal of Applied Physics}, 128(18).

\bibitem[Atzeni et~al., 2004]{atzeni2004physics}
Atzeni, S., Meyer Ter~Vehn, J., et~al. (2004).
\newblock {\em The physics of inertial fusion-Hydrodynamics, dense plasma
  physics, beam-plasma interaction}.
\newblock Clarendon Press-Oxford University Press.

\bibitem[Benedict et~al., 2014]{benedict2014multiphase}
Benedict, L.~X., Driver, K.~P., Hamel, S., Militzer, B., Qi, T., Correa, A.~A.,
  Saul, A., and Schwegler, E. (2014).
\newblock Multiphase equation of state for carbon addressing high pressures and
  temperatures.
\newblock {\em Physical Review B}, 89(22):224109.

\bibitem[Chen et~al., 2008]{chen2008dissociation}
Chen, Q., Cai, L., Zhang, Y., and Gu, Y. (2008).
\newblock The dissociation and equation of state of dense fluid oxygen at high
  pressures and high temperatures.
\newblock {\em The Journal of Chemical Physics}, 128(10).

\bibitem[Chen et~al., 2006]{chen2006self}
Chen, Q.-F., Cai, L.-C., Zhang, Y., Gu, Y.-J., and Jing, F.-Q. (2006).
\newblock Self-consistent fluid variational theory for the dissociation of
  dense nitrogen.
\newblock {\em The Journal of chemical physics}, 124(7).

\bibitem[Chen et~al., 2021]{chen2021divide}
Chen, X., Cheng, J.~Q., and Xie, M.-g. (2021).
\newblock Divide-and-conquer methods for big data analysis.
\newblock {\em arXiv preprint arXiv:2102.10771}.

\bibitem[Cox, 2018]{cox2018generating}
Cox, G. (2018).
\newblock Generating a multiphase equation of state with swarm intelligence.
\newblock In {\em AIP Conference Proceedings}, volume 1979, page 040002. AIP
  Publishing LLC.

\bibitem[Cox and Christie, 2015]{cox2015swarm}
Cox, G.~A. and Christie, M.~A. (2015).
\newblock Fitting of a multiphase equation of state with swarm intelligence.
\newblock {\em Journal of Physics: Condensed Matter}, 27:405201.

\bibitem[Curry and Schoenberg, 1988]{curry1988polya}
Curry, H.~B. and Schoenberg, I.~J. (1988).
\newblock On p{\'o}lya frequency functions iv: the fundamental spline functions
  and their limits.
\newblock In {\em IJ Schoenberg selected papers}, pages 347--383. Springer.

\bibitem[Datta et~al., 2016]{datta2016hierarchical}
Datta, A., Banerjee, S., Finley, A.~O., and Gelfand, A.~E. (2016).
\newblock Hierarchical nearest-neighbor gaussian process models for large
  geostatistical datasets.
\newblock {\em Journal of the American Statistical Association},
  111(514):800--812.

\bibitem[Drake, 2010]{drake2010high}
Drake, R.~P. (2010).
\newblock High-energy-density physics.
\newblock {\em Physics Today}, 63(6):28--33.

\bibitem[Efron and Tibshirani, 1994]{efron1994introduction}
Efron, B. and Tibshirani, R.~J. (1994).
\newblock {\em An Introduction to the Bootstrap}.
\newblock Chapman and Hall/CRC.

\bibitem[Gneiting and Raftery, 2007]{gneiting2007strictly}
Gneiting, T. and Raftery, A.~E. (2007).
\newblock Strictly proper scoring rules, prediction, and estimation.
\newblock {\em Journal of the American Statistical Association},
  102(477):359--378.

\bibitem[Gramacy and Lee, 2008]{gramacy2008bayesian}
Gramacy, R.~B. and Lee, H. K.~H. (2008).
\newblock Bayesian treed gaussian process models with an application to
  computer modeling.
\newblock {\em Journal of the American Statistical Association},
  103(483):1119--1130.

\bibitem[Khosravi et~al., 2010]{khosravi2010lower}
Khosravi, A., Nahavandi, S., Creighton, D., and Atiya, A.~F. (2010).
\newblock Lower upper bound estimation method for construction of neural
  network-based prediction intervals.
\newblock {\em IEEE transactions on neural networks}, 22(3):337--346.

\bibitem[Kim et~al., 2005]{KimMallickHolmes2005}
Kim, H.-M., Mallick, B.~K., and Holmes, C.~C. (2005).
\newblock Analyzing nonstationary spatial data using piecewise gaussian
  processes.
\newblock {\em Journal of the American Statistical Association},
  100(470):653--668.

\bibitem[Kritcher et~al., 2022]{kritcher2022design}
Kritcher, A., Young, C., Robey, H., Weber, C., Zylstra, A., Hurricane, O.,
  Callahan, D., Ralph, J., Ross, J., Baker, K., et~al. (2022).
\newblock Design of inertial fusion implosions reaching the burning plasma
  regime.
\newblock {\em Nature Physics}, 18(3):251--258.

\bibitem[Landau and Lifshitz, 1980]{landau1980statistical}
Landau, L.~D. and Lifshitz, E.~M. (1980).
\newblock {\em Statistical Physics}, volume~5 of {\em Course of Theoretical
  Physics}.
\newblock Pergamon Press, 3 edition.

\bibitem[Lee and Park, 2023]{lee2023scalable}
Lee, B.~S. and Park, J. (2023).
\newblock A scalable partitioned approach to model massive nonstationary
  non-gaussian spatial datasets.
\newblock {\em Technometrics}, 65(1):105--116.

\bibitem[Lindemann, 1910]{lindemann1910calculation}
Lindemann, F. (1910).
\newblock The calculation of molecular vibration frequency.
\newblock {\em Z. phys}, 11:609--612.

\bibitem[Lindgren et~al., 2011]{lindgren2011explicit}
Lindgren, F., Rue, H., and Lindstr{\"o}m, J. (2011).
\newblock An explicit link between gaussian fields and gaussian markov random
  fields: the stochastic partial differential equation approach.
\newblock {\em Journal of the royal statistical society: Series b (statistical
  methodology)}, 73(4):423--498.

\bibitem[Lindl et~al., 2004]{lindl2004physics}
Lindl, J.~D., Amendt, P., Berger, R.~L., Glendinning, S.~G., Glenzer, S.~H.,
  Haan, S.~W., Kauffman, R.~L., Landen, O.~L., and Suter, L.~J. (2004).
\newblock The physics basis for ignition using indirect-drive targets on the
  national ignition facility.
\newblock {\em Physics of Plasmas}, 11(2):339--491.

\bibitem[Lindquist and Jadrich, 2022]{lindquist2022uncertainty}
Lindquist, B.~A. and Jadrich, R.~B. (2022).
\newblock Uncertainty quantification for a multi-phase carbon equation of state
  model.
\newblock {\em Journal of Applied Physics}, 131(15).

\bibitem[Lindquist and Jadrich, 2023]{lindquist2023uncertainty}
Lindquist, B.~A. and Jadrich, R.~B. (2023).
\newblock Uncertainty quantification for a multi-phase magnesium equation of
  state.
\newblock In {\em AIP conference proceedings}, volume 2844, page 310006. AIP
  Publishing LLC.

\bibitem[Lyon et~al., 1992]{lyon1992alamos}
Lyon, S.~P., Johnson, J.~D., and {Group T-1} (1992).
\newblock {SESAME}: The {Los Alamos National Laboratory} equation-of-state
  database.
\newblock Technical Report LA-UR-92-3407, Los Alamos National Laboratory, Los
  Alamos, NM.

\bibitem[Maerzke et~al., 2019]{maerzke2019equations}
Maerzke, K.~A., Coe, J.~D., Ticknor, C., Leiding, J.~A., Tinka~Gammel, J., and
  Welch, C.~F. (2019).
\newblock Equations of state for polyethylene and its shock-driven
  decomposition products.
\newblock {\em Journal of Applied Physics}, 126(4).

\bibitem[Mentzer and Peterson, 2023]{mentzer2023neural}
Mentzer, K.~L. and Peterson, J.~L. (2023).
\newblock Neural network surrogate models for equations of state.
\newblock {\em Physics of Plasmas}, 30(3).

\bibitem[Meyer, 2008]{meyer2008inference}
Meyer, M.~C. (2008).
\newblock Inference using shape-restricted regression splines.
\newblock {\em The Annals of Applied Statistics}, 2(3):1013--1033.

\bibitem[More et~al., 1988]{more1988new}
More, R.~M., Warren, K., Young, D., and Zimmerman, G. (1988).
\newblock A new quotidian equation of state (qeos) for hot dense matter.
\newblock {\em The Physics of fluids}, 31(10):3059--3078.

\bibitem[Myint et~al., 2017]{myint2017free}
Myint, P.~C., Benedict, L.~X., and Belof, J.~L. (2017).
\newblock Free energy models for ice vii and liquid water derived from
  pressure, entropy, and heat capacity relations.
\newblock {\em The Journal of Chemical Physics}, 147(8).

\bibitem[Myint et~al., 2023]{myint2023fluid}
Myint, P.~C., Root, S., Wu, C.~J., Clay, R.~C., Lopez, A., Hanshaw, H.~L.,
  Lemke, R.~W., Bliss, D.~E., Hanson, D.~L., Flicker, D.~G., et~al. (2023).
\newblock Fluid-phase helium: Shock-compression experiments, quantum molecular
  dynamics simulations, and development of an equation of state.
\newblock {\em Physical Review B}, 108(13):134202.

\bibitem[Paciorek and Schervish, 2006]{paciorek2006spatial}
Paciorek, C.~J. and Schervish, M.~J. (2006).
\newblock Spatial modelling using a new class of nonstationary covariance
  functions.
\newblock {\em Environmetrics: The official journal of the International
  Environmetrics Society}, 17(5):483--506.

\bibitem[Pearce et~al., 2018]{pearce2018high}
Pearce, T., Brintrup, A., Zaki, M., and Neely, A. (2018).
\newblock High-quality prediction intervals for deep learning: A
  distribution-free, ensembled approach.
\newblock In {\em International conference on machine learning}, pages
  4075--4084. PMLR.

\bibitem[Ramsay, 1988]{ramsay1988monotone}
Ramsay, J.~O. (1988).
\newblock Monotone regression splines in action.
\newblock {\em Statistical science}, pages 425--441.

\bibitem[Rasmussen, 2003]{rasmussen2003gaussian}
Rasmussen, C.~E. (2003).
\newblock Gaussian processes in machine learning.
\newblock In {\em Summer school on machine learning}, pages 63--71. Springer.

\bibitem[Rehn et~al., 2021]{rehn2021multiphase}
Rehn, D.~A., Greeff, C.~W., Burakovsky, L., Sheppard, D.~G., and Crockett,
  S.~D. (2021).
\newblock Multiphase tin equation of state using density functional theory.
\newblock {\em Physical Review B}, 103(18):184102.

\bibitem[Ross, 1985]{ross1985matter}
Ross, M. (1985).
\newblock Matter under extreme conditions of temperature and pressure.
\newblock {\em Reports on Progress in Physics}, 48(1):1--52.

\bibitem[Ruppert et~al., 2003]{ruppert2003semiparametric}
Ruppert, D., Wand, M.~P., Carroll, R.~J., and Ruppert, D. (2003).
\newblock {\em Semiparametric regression}, volume~12.
\newblock Cambridge university press Cambridge.

\bibitem[Sampson and Guttorp, 1992]{sampson1992nonparametric}
Sampson, P.~D. and Guttorp, P. (1992).
\newblock Nonparametric estimation of nonstationary spatial covariance
  structure.
\newblock {\em Journal of the American Statistical Association},
  87(417):108--119.

\bibitem[Schmid and Andrews, 2025]{schmid2025simultaneous}
Schmid, A.~C. and Andrews, S.~A. (2025).
\newblock Simultaneous inference of equation of state parameters and unknown
  data errors with uncertainty quantification via hierarchical bayesian
  posterior maximization.
\newblock {\em Journal of Applied Physics}, 138(15).

\bibitem[Schmidt and Guttorp, 2020]{schmidt2020flexible}
Schmidt, A.~M. and Guttorp, P. (2020).
\newblock Flexible spatial covariance functions.
\newblock {\em Spatial Statistics}, 37:100416.

\bibitem[Seto et~al., 2008]{seto2008fate}
Seto, Y., Hamane, D., Nagai, T., and Fujino, K. (2008).
\newblock Fate of carbonates within oceanic plates subducted to the lower
  mantle, and a possible mechanism of diamond formation.
\newblock {\em Physics and Chemistry of Minerals}, 35(4):223--229.

\bibitem[Sharma et~al., 2024]{sharma2024learning}
Sharma, H., Gaffney, J.~A., Tsapetis, D., and Shields, M.~D. (2024).
\newblock Learning thermodynamically constrained equations of state with
  uncertainty.
\newblock {\em APL Machine Learning}, 2(1).

\bibitem[Stacey, 2005]{stacey2005high}
Stacey, F.~D. (2005).
\newblock High pressure equations of state and planetary interiors.
\newblock {\em Reports on progress in physics}, 68(2):341--383.

\bibitem[Toro, 2009]{toro2009equations}
Toro, E.~F. (2009).
\newblock The equations of fluid dynamics.
\newblock In {\em Riemann Solvers and Numerical Methods for Fluid Dynamics: A
  Practical Introduction}, pages 1--40. Springer.

\bibitem[van Thiel and Ree, 1987]{van1987properties}
van Thiel, M. and Ree, F.~H. (1987).
\newblock Properties of carbon clusters in tnt detonation products:
  Graphite-diamond transition.
\newblock {\em Journal of Applied Physics}, 62(5):1761--1767.

\bibitem[Vecchia, 1988]{vecchia1988estimation}
Vecchia, A.~V. (1988).
\newblock Estimation and model identification for continuous spatial processes.
\newblock {\em Journal of the Royal Statistical Society Series B: Statistical
  Methodology}, 50(2):297--312.

\bibitem[Wahba, 1990]{wahba1990spline}
Wahba, G. (1990).
\newblock {\em Spline Models for Observational Data}.
\newblock SIAM.

\bibitem[Wang and Yan, 2021]{wang2021shape}
Wang, W. and Yan, J. (2021).
\newblock Shape-restricted regression splines with {R} package
  \texttt{splines2}.
\newblock {\em Journal of Data Science}, 19(3):498--517.

\bibitem[Wood, 2017]{wood2017generalized}
Wood, S.~N. (2017).
\newblock {\em Generalized Additive Models: An Introduction with R}.
\newblock Chapman and Hall/CRC.

\bibitem[Wu et~al., 2023]{wu2023wide}
Wu, C.~J., Benedict, L.~X., Myint, P.~C., Hamel, S., Prisbrey, C.~J., and Leek,
  J.~R. (2023).
\newblock Wide-ranged multiphase equation of state for iron and model
  variations addressing uncertainties in high-pressure melting.
\newblock {\em Physical Review B}, 108(1):014102.

\bibitem[Wu et~al., 2019]{wu2019equation}
Wu, C.~J., Young, D.~A., Sterne, P.~A., and Myint, P.~C. (2019).
\newblock Equation of state for a chemically dissociative, polyatomic system:
  Carbon dioxide.
\newblock {\em The Journal of Chemical Physics}, 151(22).

\bibitem[Xu and Reich, 2023]{xu2023bayesian}
Xu, S.~G. and Reich, B.~J. (2023).
\newblock Bayesian nonparametric quantile process regression and estimation of
  marginal quantile effects.
\newblock {\em Biometrics}, 79(1):151--164.

\bibitem[Zhang et~al., 2015]{Zhang2015}
Zhang, Y., Duchi, J., and Wainwright, M. (2015).
\newblock Divide and conquer kernel ridge regression.
\newblock {\em Journal of Machine Learning Research}, 16:3299--3340.

\end{thebibliography}


\begin{thebibliography}{}

\bibitem[Efron and Tibshirani, 1994]{efron1994introduction}
Efron, B. and Tibshirani, R.~J. (1994).
\newblock {\em An Introduction to the Bootstrap}.
\newblock Chapman and Hall/CRC.

\bibitem[Gneiting and Raftery, 2007]{gneiting2007strictly}
Gneiting, T. and Raftery, A.~E. (2007).
\newblock Strictly proper scoring rules, prediction, and estimation.
\newblock {\em Journal of the American Statistical Association},
  102(477):359--378.

\bibitem[Khosravi et~al., 2010]{khosravi2010lower}
Khosravi, A., Nahavandi, S., Creighton, D., and Atiya, A.~F. (2010).
\newblock Lower upper bound estimation method for construction of neural
  network-based prediction intervals.
\newblock {\em IEEE transactions on neural networks}, 22(3):337--346.

\bibitem[Pearce et~al., 2018]{pearce2018high}
Pearce, T., Brintrup, A., Zaki, M., and Neely, A. (2018).
\newblock High-quality prediction intervals for deep learning: A
  distribution-free, ensembled approach.
\newblock In {\em International conference on machine learning}, pages
  4075--4084. PMLR.

\end{thebibliography}
